\documentclass[usenatbib]{mn2e}
\usepackage{times}
\usepackage{epsfig}

\def\mh1{$M_{\rm H_{I}}$}

\def\etal{{\it et al.}\thinspace}

\title[]{IGM Heating in Fossil Galaxy Groups}
\author [H. Miraghaei, et al]{H. Miraghaei$^{1,}$$^2$, H. G. Khosroshahi$^1$, H.-R. Kl\"ockner $^{3,}$$^4$, T. J. Ponman$^5$, 
  N. N. Jetha$^6$, \newauthor S. Raychaudhury$^{5}$ \\
$^{1}$School of Astronomy, Institute for Research in Fundamental Sciences, PO Box 19395-5531, Tehran, Iran\\
$^{2}$Department of Physics, Sharif University of Technology PO Box 11365-9161, Tehran, Iran\\
$^{3}$Subdepartment of Astrophysics, University of Oxford, Denys-Wilkinson Building, Keble Road,
Oxford, OX1 3RH, UK\\
$^{4}$Max-Planck-Institut f\"ur Radioastronomie, Auf dem H\"ugel 69, 53121 Bonn, Germany\\
$^{5}$School of Physics and Astronomy, University of Birmingham, Edgbaston, Birmingham B15 2TT, UK\\
$^{6}$Center for Space Plasma and Aeronomic Research (CSPAR), University of Alabama in Huntsville, Huntsville, AL 35805, USA\\
}
\begin{document}
\date{}
\pagerange{\pageref{firstpage}--\pageref{lastpage}} \pubyear{}

\maketitle

\label{firstpage}
\begin{abstract}

We study intergalactic medium (IGM) heating in a sample of five fossil galaxy groups by using their radio properties at 610 MHz and 1.4 GHz. The power by radio jets introducing mechanical heating for the sampled objects is not sufficient enough to suppress the cooling flow. Therefore, we discussed shock-, vortex heating, and conduction as alternative heating processes. Further, the 1.4 GHz and 610 MHz radio luminosities of fossil groups are compared to a sample of normal galaxy groups of the same radio brightest (BGGs), stellar mass, and total group stellar mass, quantified using the $K$-band luminosity. It appears that the fossil BGGs are under luminous at 1.4 GHz and 610 MHz for a given BGG stellar mass and luminosity, in comparison to a general population of the groups. In addition, we explore how the bolometric radio luminosity of fossil sample depends on clusters and groups characteristics. Using the HIghest X-ray FLUx Galaxy Cluster Sample (HIFLUGCS) as a control sample we found that the large-scale behaviours of fossil galaxy groups are consistent with their relaxed and virialised nature.  

\end{abstract}

\section{Introduction}
\label{sec:Intro}

Galaxy groups and clusters are known as the most massive gravitationally bound systems in the Universe. Their gravitational potential is sufficient enough to bound large amounts of intergalactic gas within its central region. This intergalactic medium (IGM) shows core temperatures of the order of million of Kelvin, which has been revealed by X-ray observations, as one of the key properties of these structures. To understand the observed state of the gas, various heating- and cooling-processes have been proposed to work and are related to the thermal history in the formation of the group or cluster. To study the interplay of these processes fossil galaxy groups seem to be promising candidates.\\

Fossil groups are dominated by a giant elliptical galaxy at the centroid of their X-ray emission and have been discovered in the mid 1990 using selection criteria in the optical and X-ray bands \citep{Ponman94}. The morphology of the luminous X-ray emission in fossils is regular and symmetric, indicating the absence of a recent group-scale merger. \cite{kmpj06,kpj07} reported a higher dark matter concentration in fossil halos compared to non-fossil groups and clusters with similar masses. In addition, theoretical studies using the Millennium simulations showed that most of the halo mass in fossil galaxy groups is assembled earlier compared to the general population of galaxy groups \citep{ dariush07, dariush10}.
 All this together and the absent of $L_{\star}$ galaxies suggest that they arguably are old galaxy groups.
Up to now more than one hundred galaxy groups have been classified as "fossil groups" by different studies \citep{ Ponman94, Jones03,
Yoshioka04, kjp04, Vikhlinin99, Voevodkin10, Santos07,  Schirmer10, Barbera09, Lopes10, Proctor11, Pierini11, Miller12, Eigenthaler09}. However, the selection criteria are not strictly met in all studies and vary from the convention introduced by \citet{Jones03}. According to the \cite{Jones03} definition, observationally a galaxy group is fossil if it has an X-ray luminosity of $L_{X,bol}\geq {10}^{42} {h_{50}}^{-2}$ erg {s}$^{-1}$ and its dominant galaxy is at least two magnitudes brighter, in $R$-band, than the second ranked galaxy within half the projected virial radius of the group ($\Delta m_{12}>2$). Note that an alternative criteria with $\Delta m_{14}>2.5$ within the same radius, reveals fifty percent more early-formed systems \citep{dariush10}.

Generally, the IGM in group and cluster of galaxies reaches very hot core temperatures and therefore emits X-rays in form of bremsstrahlung and atomic line emission. The cooling process itself works via the energy loss through both continuum- and spectral-emission and cools down the inter-galactic gas such that the gas is falling deeper into the potential well. In this process the gas density increases at the core and further boosts the X-ray emission, as the X-ray emissivity is proportional to the square of the gas density \citep{Sarazin88}. This cooling flow will produce a strong surface brightness peak toward the centre of clusters and a drop in temperature in the core \citep{Peterson06}. This classical picture does not consider any heating mechanism, which must be in place, because X-ray observations of galaxy clusters by Chandra and XMM-Newton show no sign of gas temperatures below ${\sim}$ keV. Among all the heating sources suggested in the literature, AGNs are the most promising candidates to be powerful enough to heat up the ICM above the observed gas temperatures \citep{Gitti12}. In particular several mechanism are proposed:  (I) The AGN can heat up the IGM by depositing relativistic electrons through jets and with the mechanical work of expanding bubbles. The dissipation heat of shock waves and sound waves induced by AGNs are also considered (see \cite{McNamara07}. (II) Cosmic rays could heat via AGN activity in addition to mechanical heating has been proposed \citep{Guoh08}. 
(III) Conduction of heat from hot outer layers of cluster into the centre is also suggested \citep{Ruszkowski02, Ruszkowski11, Bregman88}. (IV) Galaxy merger 
\citep{Gomez02, Sarazin02, Kempner03, Motl04} and intra-cluster supernovae \citep{Domainko04} would provide sufficient energy to be injected to the ICM.

 Apart from the heating scenario the efficiency of mechanical heating to quench the cooling rate is also investigated in the literature. \cite{ Birzan04, DunnF06} showed that the bubble enthalpy is sufficient to balance the cooling at least for half of the galaxy clusters in their samples. The efficiency to quench the cooling rate does dramatically increase in elliptical galaxies as \cite{Nulsen07} showed for a sample of nearby giant elliptical galaxies. The energy released by AGN jets in the form of bubbles are estimated by several authors \citep{Nusser06, Begelman01}. Finding these cavities in the X-ray surface brightness distribution of galaxy clusters by Chandra and XMM-Newton would support the picture of bubble heating in the ICM. In many of these cavities the radio emissions at Gigahertz or above might fall below the detection threshold and thus low radio frequency observations may enables us to trace these cavities \citep{Jetha08}.

The IGM heating due to AGN is generally blamed for the observed discrepancies in the scaling relations in the galaxy groups regime \citep{McCarthy10}, such as the deviations from the self-similar scaling observed for galaxy clusters \citep{Helsdon00, Osmond04}.  However, due to the complex nature of the galaxy groups, which are subject to  both group scale mergers and galaxy-galaxy mergers, the share of the AGN in the IGM heating can not be estimated accurately.  Having a sample of galaxy groups in which there are no signs of recent galaxy mergers within the group and the groups scale mergers is very attractive as it eliminates some of the above complexities. As mentioned earlier, fossil galaxy groups offer both these advantages at the same time. The morphological studies of the central galaxy in fossil groups, the isophotal shapes in particular, shows that they are not boxy type \cite{kpj06,Smith10}, which points at the lack of dry merger in their recent evolutionary history. Furthermore, the formation history of the groups also point to an early formation epoch. There are, however, counter evidences for this, presented by La Barbera et al (2012).

In this study we aim to investigate the heating mechanism in place of the intercluster medium in fossil galaxy groups. For this, we observed a sample of well-known fossil group in the radio regime with the Giant Metrewave Radio Telescope (GMRT) at 610 MHz and 1.4 GHz. In addition, the optical and X-ray properties are used to investigate the characteristics of this sample of fossil groups in comparision with a sample of normal galaxy groups and clusters in radio band.  This is the first study to focus on the IGM heating in a sample of fossil galaxy groups.

In section \ref{sec:Sample}, the sample and the radio observations are discussed including the data reduction. 
The radio properties are shown in section \ref{sec:Result} combined with a brief description on the X-ray and optical characteristics. In section \ref{sec:Heating} the heating mechanism and their energy requirements are discussed. In section \ref{sec:Accretion} we consider feeding mechanisms of the central AGN and section \ref{sec:stat} investigates the difference between fossil and non-fossil groups. For this simple statistical study, radio luminosity and other properties of galaxy groups and clusters are used. The summery and conclusion is presented in section \ref{sec:summery}.  Throughout this paper we assumed a $\Lambda CDM$ cosmology with the following parameters: $\Omega_m=0.27$, $\Omega_\Lambda=0.73$ and $H_0=100 h$ km s$^{-1}$ Mpc$^{-1}$ where $h$ = 0.71.\\

\begin{table*}
\begin{center}
\caption{Fossil groups sample 
\label{table1}
}
\begin{tabular}{llllllllll}
\hline
Object & $Ra$ & $Dec$ &   $z$ &  $M_{r}$ & ${M_{k}}^a$& ${M_{k}}^b$& $L_{x}$& ${\Delta m}^c$&  \\
 \hline 

RX J1331.5+1108 (J1331)&13:31:30.2&+11:08:04&0.081&-22.43& -24.23 &-25.01& 2.1 E42  &2.0\\
RX J1416.4+2315 (J1416)&14:16:26.9&+23:15:32&0.137&-23.84& -25.33 &-26.47& 1.7 E44 &2.4\\
RX J1552.2+2013 (J1552)&15:52:12.5&+20:13:32&0.135&-23.64&  -25.71 & -26.30&  6 E43  &2.3\\
NGC 6482&17:51:48.8&+23:04:19&0.013&-21.36& -22.91 &-25.28& 1.11 E42 &2.06\\
ESO 3060170 &05:40:06.7&-40:50:11&0.036&-23.66& -23.74&-25.98 & 6.6 E43&2.61\\
\hline 
(a)  from 2MASS-PSC \\
(b)  from 2MASS-XSC \\
(c)  luminosity Gap\\
\end{tabular}

\end{center}
\end{table*}

\section {Sample, Radio Observations and Data Reduction }
\label{sec:Sample}
In order to investigate the source of heating in fossil group, our sample of 5 fossil systems has been selected on the basis of the flux-limited sample of fossil groups reported in
\cite{ kpj07,Jones03}. Due to the sensitivity limitation of our radio observations the most distant fossil groups have been excluded and therefore reduced the sample from 7 to 5 systems. The sample varies in halo mass, with the NGC6482 (3.5${\times}10^{12}$ M$_{\odot}$) \citep{kjp04} being the least massive and the  RX~J1416.4+2315 (179${\times}10^{12}$ M$_{\odot}$) being the most massive \citep{kmpj06}, and offers a reasonable dynamical range to investigate the heating mechanism caused by the AGN.

Based on optical and X-ray properties, fossil groups are considered to reside between normal groups and clusters of galaxies \citep{kpj07,Hess12}. In order to explore the group-scaling behaviour of such fossil and non-fossil systems in relation to the AGN activities, various reference samples have been used. In particular, for a statistical comparison in section~\ref{sec:stat} we used a non-fossil group sample observed at 610~MHz by \cite{Giacintucci11} and the HIFLUGCS galaxy cluster sample observed at X-ray \citep{Reiprich02}.
In order to investigate the workings of the mechanical heating discussed in section~\ref{sec:Heating1} we used a galaxy cluster sample as a reference, with mechanical luminosity reported in \cite{Birzan04}.

The fossil group sample is listed in Table 1. The $K$-band absolute magnitudes have been obtained from 2MASS \footnote{Two Micron All Sky Survey.}.
Radio observations of all sources were performed by the GMRT.  The observations for the first three sources listed in Table 2, were carried out during 10-21 July 2009.  We included archive data for NGC 6482 and RX J1416.4+2315 observed in 2006 \citep{miraghaei12}.
Observing frequencies were chosen to be at 1.4 GHz and 610 MHz with 16 MHz band width in both lower-side band and upper-side band.  Except for NGC~6482 the 1.4 GHz information was taken from the VLA archive.  
The entire data bandwidth was divided into 128 channels with the spectral resolution  of 125 kHz. Total integration time for each source was  $\sim$10 hours in a single or in two separate slots.  Each observation started with a flux calibrator run (20-30 minutes), 5 minutes for the selected phase calibrator, and followed by 30 minutes on the target source. After that the observation run periodically switched between the phase calibrator and the target source, until it ended with a flux calibrator observation. For observations lasting longer than 9~hours, we observed a flux calibrator at the middle of observing run. The minimum and the maximum baselines are ${\sim}$ 0.4--100 k${\lambda}$ for the 1.4 GHz observations and ${\sim}$ 0.2--50 k${\lambda}$ for the 610 MHz observations and the resulting angular resolutions are quoted in Table 4. Data sets are record in Stokes RR and LL and therefore Stokes I maps have been produced for the sources. The physical extend of the fossil group sample is estimated by its virial radius and varies from 7 to 43~arcmin (Table 3) at their given redshifts. The full width at half maximum of the primary beam of the GMRT is 26 arcmin and 44 arcmin at 1.4 GHz and 610 MHz, respectively. Therefore, the field of view is sufficiently to observe the entire extend of the groups except for the group ESO~3060170 at 1.4 GHz.  Thus we can identify group members and other radio structures of the IGM \citep{Feretti12} as well as their BGGs emissions.

\begin{table*}
\begin{center}
\caption{The radio observations of the sample using the GMRT at 1.4 GHz and 610 MHz.
(a) Angular separation between the the source and the selected phase calibrator.
\label{table2}
}
\begin{tabular}{lrrccccc}
\hline
Object&$\nu$&Observation&$t$&Flux&Phase& {Separation }$^a$&\\
&[MHz]&date&[h]&Calibrator&Calibrator&[deg]&\\
\hline 

RX J1331.5+1108&1280&10-July-2009&6&3C286&1347+122&4&\\
RX J1331.5+1108&601&17,18-July-2009&2.5&3C147, 3C286&1330+251&14&\\
RX J1416.4+2315&1280&4,5-July-2005&14&3C286, 3C287&1407+284&6\\
RX J1416.4+2315&601&12-July-2005&14&3C286, 3C287&1407+284&6\\
RX J1552.2+2013&1280&13-July-2009&6.5&3C286, 3C48&1609+266&7\\
RX J1552.2+2013&601&20-July-2009&5.5&3C286, 3C48&1419+064, 1609+266&26, 7\\
NGC 6482&618&21-DEC-2006&6&3C286, 3C468.1&1609+266&24\\
ESO 3060170&1280&13-July-2009&6&3C147, 3C48&0538- 440, 0440- 435&3, 11\\
ESO 3060170&601&18,19-July-2009&2.5&3C147, 3C48&0521- 207&20\\
\hline  
\end{tabular}
\end{center}
\end{table*}

\begin{table*}
\begin{center}
\caption{Some basic characteristics of the fossil groups sample.   
\label{table3}
}
\begin{tabular}{lllllllr}
\hline
object & $z$ & kpc {arcsec}$^{-1}$ &   $\sigma$ &  $R_{virial}$ &angular size   \\
  &  &  &  [km {s}$^{-1}$] &  [Mpc] &[arcmin]  \\
 
\hline 

RX J1331.5+1108&0.081&1.53&236$\pm$79&0.674$\pm$0.225&7.34$\pm$2\\
RX J1416.4+2315&0.137&2.44&694$\pm$120&1.982$\pm$0.342&13.53$\pm$2\\
RX J1552.2+2013&0.135&2.40&721$\pm$150&2.06$\pm$0.428&14.30$\pm$3\\
NGC 6482&0.013&0.27&115$\pm$38&0.325$\pm$0.108&20.06$\pm$7\\
ESO 3060170&0.036&0.71&648$\pm$160&1.851$\pm$0.457&43.45$\pm$10\\
\hline  
\end{tabular}
\end{center}
\end{table*}

\smallskip

Data reduction was carried out in AIPS (Astronomical Image Processing System). We followed the standard procedure for the calibration, by calibrating the flux and phase calibrators and applying the solutions to the target sources. For bandpass calibration we used the flux calibrator, after applying the bandpass the spectrum of 128 channels has been averaged into 10 channels.
The used flux- and phase-calibrators are listed in Table 2. The separation of the target sources and their corresponding phase calibrators varies from 3 to 26 degrees. 

For imaging, we initially produced a low quality image with one facet to check the quality of the calibration. Most of the primary images suffered from imperfect flagging and/or poor calibration. We used the task {\tt FLGIT} to remove radio frequency interference (RFI), however, the effect of poorly phase calibrated baselines was found to be significant. The only solution to distinguish and remove bad baselines, and eventually decrease the rms of the images, was to image each baseline individually to identify and flag the RFI.
A routine developed by N. Kantharia and R. Nityananda (private communication) was used to find bad baselines specially those with high flux density values.  We also used an earlier version of the pipeline developed by H.-R. Kl\"{o}ckner (e.g. \cite{Mauch13}), which could find and remove most of the bad baselines and improve the quality of the images.

For imaging, we used the task {\tt IMAGR} with multi faceting option in both frequencies to cover the primary beam and correct for the W-term effect of bright sources with a large separation angle with respect to the field centre. During the cleaning, we carefully identified the suspected sources to prevent errors possibly created within the self-calibration process. The dataset has been self-calibrated in phase to further improve the quality of the image. The entire field of view toward the target source as been finally imaged by combining the facets using the task {\tt FLATN} in AIPS.  Although the noise level was very high around some strong point sources, the self-calibration improved the quality of the images, substantially. When observations were split into two runs, a self-calibration sequence was applied after combining the UV data by using the task {\tt DBCON} in AIPS. In the case of detecting a radio extension around the central source, different UV weighting was used in order to reach the best noise performance (AIPS input of robust 0). We reduced both the LSB and the USB data but finally chose the best produced map among them.

In particular for NGC 6482, the flux density calibrators at 610 MHz, were 3C286 and 3C468.1 at the start and the end of observations, respectively.  In the absence of prior flux density measurements for the 3C468.1 at 610 MHz, we used the 750 MHz flux density reported in NED \footnote{Nasa Extragalactic Database},  $f$  = 9.7 mJy, and the spectral index of $\alpha$ = --\.0.728 ($f {\propto} {\nu}^{\alpha}$) reported in \cite{Kellermann69} to estimate the flux density. Using this method the estimated flux density for 3C468.1 at 610 MHz is 9.06~mJy, which has been used within the calibration process.
\section {Results of GMRT observations}
\label{sec:Result}

Radio images are shown in Figs.~\ref{fig1}-\ref{fig10} with the angular resolution of 2-10 arcseconds.  
The noise level achieved in the 1.4~GHz maps, ranges from 50 $\mu$Jy  to 120 $\mu$Jy  and in 610 MHz maps ranges from 
120 $\mu$Jy  to 160 $\mu$Jy. The dynamic range in cleaned images varies from 300 to 1000 in both frequencies. Table 4 lists the flux density measurements of the sources. The flux density and extend of the sources were calculated by fitting a Gaussian function to the image using the AIPS task {\tt JMFIT}. Note that all target sources have been observed at the phase centre and therefore the resulting flux density estimates are not affected by a potential error of the primary beam model of the GMRT. The noise level around the source is also reported. Generally, the noise level over the entire map is lower than noise level around the sources. We applied a 3$\sigma$ detection limit for a reliable source detection. The spectral indices are obtained according to the relation $S {\propto} {\nu}^{\alpha}$, within the two observed frequencies. The synthesized beam is determined by the robust weighting scheme of 0 and is illustrated in all the radio images. We also listed the flux densities measures of our sample at 1.4 GHz from the NVSS \footnote{NRAO VLA Sky Survey} or the FIRST \footnote{Faint Images of the Radio Sky at Twenty-Centimeters} catalogues in order to compare these measures with our results. We also reported radio flux densities of objects from PMN \footnote{Parkes-MIT-NRAO} and SUMSS \footnote{Sydney University Molonglo Sky Survey} catalogues in the case of existences (Table \ref{table5}). The calculated radio luminosities, resolutions and objects sizes in both frequencies are also presented in Table 4. In this study, we assumed 10 per cent amplitude calibration errors for GMRT data.

For all objects a radio emission has been detected which coincides with the corresponding BGG in the optical band and the peak brightness spot of the X-ray emission, except for the source J1552.2+2013. In two cases, ESO~3060170 and J1416, the radio observations show lobes extended up to 4 and 28 kpc from the central source away. In particular, the radio lobes are detected clearly in the 610 MHz observations, for both systems. But, at 1.4 GHz only  ESO 3060170 shows a radio lobe (see Figs. \ref{fig3},\ref{fig4}  and \ref{fig8},\ref{fig9}). Furthermore, all radio sources have negative spectral index around --1 ($S {\propto} {\nu}^{\alpha}$), except for J1331.5+1108. We discuss this in details in the section \ref{sec:Result1}.

\begin{table*}
\begin{center}
\caption{Results of the 1.4 GHz and 610 MHz flux densities measurements.    
\label{table4}
}
\begin{tabular}{llllllllllr}
\hline
object&$rms$&peak flux &$\nu$&flux &$\alpha$&$L_{Radio}$&$F_{1.4 GHz}$&synthesized &b$_{maj}$ $\times$ b$_{min}$\\
&&density&&density&&&&beam&\\
&$[{\mu}$Jy {beam}$^{-1}$]&[mJy {beam}$^{-1}$]&[MHz]&[mJy]&$f{\propto}{\nu}^{\alpha}$&[10$^{22}$ W {Hz}$^{-1}$]&[mJy]& [arcsec]&[arcsec]\\
 \hline 

RX J1331.5+1108&48-150&13.7&1280&14.59$\pm$1.47&+0.93$\pm$0.19&20.23&14.2$\pm$0.6$^{a}$&2.9${\times}$2.1&3.05${\times}$2.20\\
&&&&&&&12.65$^{b}$ && \\
RX J1331.5+1108&129-201&7.22&618&7.42$\pm$0.76&+0.93$\pm$0.19&10.28&-&5.7${\times}$4.5&5.5${\times}$3.6\\
RX J1416.4+2315&120&2.29&1280&2.52&-0.93&9.74&2.36$\pm$0.145$^{b}$&9.01${\times}$5.55&13${\times}$12\\
RX J1416.4+2315 A&120&3.91&601&5.54&-0.93&21.42&-&9.03${\times}$5.71&18${\times}$13\\
RX J1416.4+2315 B&120&1.04&601&1.01&${\textless}$-1.2&3.90&-&9.03${\times}$5.71&12${\times}$8\\
RX J1552.2+2013&80&-&1280&-&-&${\textless}$0.90&${\textless}$0.149$^{b}$&2.9${\times}$2.4&-\\
RX J1552.2+2013&80&-&601&-&-&${\textless}$0.90&-&7.5${\times}$6.7&-\\
NGC 6482&490&1.79&1420&1.85$\pm$0.86&-1.05$\pm$0.58&0.06&1.85$^{a}$&45${\times}$45&56${\times}$37\\
NGC 6482&122-178&3.8&616&5.38$\pm$0.56&-1.05$\pm$0.58&0.19&-&6.0${\times}$4.8&12.3${\times}$4.9\\
ESO 3060170 A&47-190&6.088&1270&8.06$\pm$0.82&-0.87$\pm$0.22&2.24&-&4.9${\times}$2.7&5.1${\times}$3.0\\
ESO 3060170 B&47-190&0.66&1270&1.34$\pm$0.23&-0.93$\pm$0.32&0.37&-&4.9${\times}$2.7&6.7${\times}$3.5\\
ESO 3060170 A&155-300&9.46&618&15.06$\pm$1.52&-0.87$\pm$0.22&4.20&-&10.9${\times}$4.2&9.5${\times}$4.6\\
ESO 3060170 B&155-300&2.23&618&2.60$\pm$0.39&-0.93$\pm$0.32&0.72&-&10.9${\times}$4.2&10.3${\times}$3.1\\

\hline 
(a)  NVSS \\
(b)  FIRST \\ 
\end{tabular}
\end{center}
\end{table*}
\subsection{RX J1331.5+1108}
\label{sec:Result1}
\begin{figure}
\center
\epsfig{file=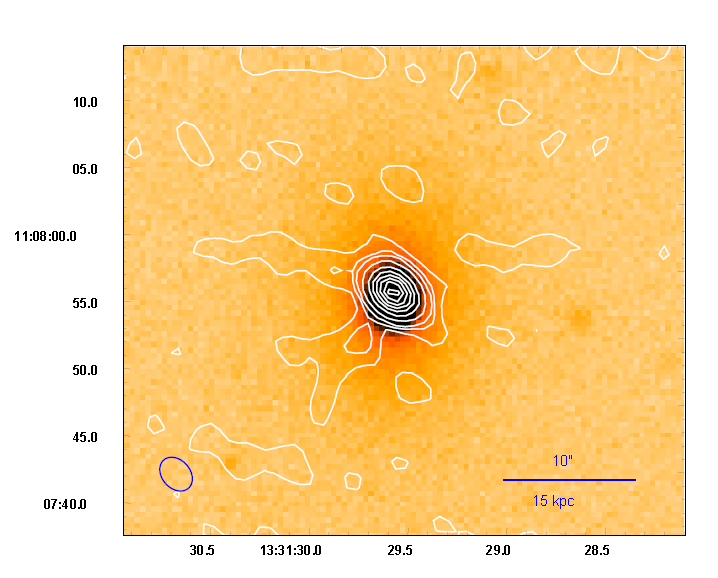 ,width=3.2in,height=2.6in}
\caption{The 1.4 GHz radio map of J1331, overlaid on SDSS $r$-band image of the central galaxy. 
Contour levels of ${\sigma}$ = 0.18 mJy ${\times}$
 2 , 3 , 4 , 5 , 10 , 20 , 30 , 40 , 50 , 60 , 70 are shown. }
\label{fig1}
\end{figure}

\begin{figure}
\center
\epsfig{file=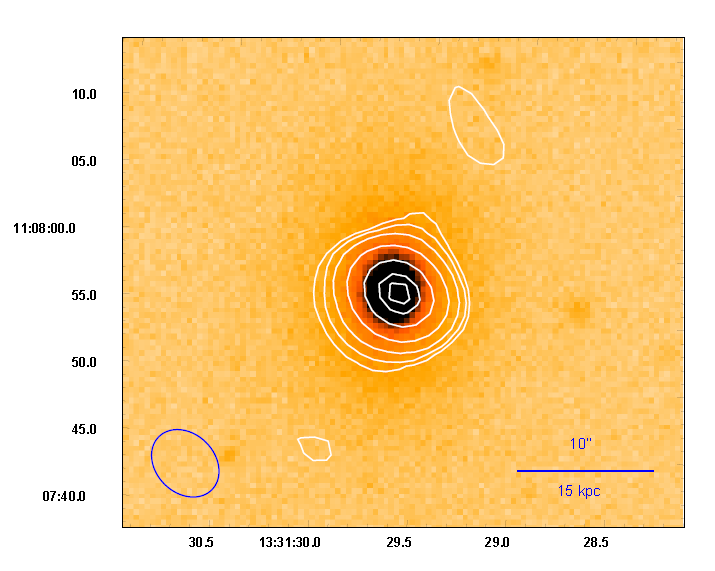,width=3.2in,height=2.6in}
\caption{The 610 MHz radio map of J1331, overlaid on SDSS $r$-band image of the central galaxy. 
Contour levels of Contour levels of ${\sigma}$ = 0.2 mJy ${\times}$
 2 , 3 , 5 , 10 , 30 , 40 , 45  are shown.}
\label{fig2}
\end{figure}

The group J1331 hosts the most the luminous radio galaxy in our sample with radio luminosity of $L_{1.4 GHz}$= 20.23${\times}10^{22}$ W {Hz}$^{-1}$, while it's relatively weak in the optical $R$-band.  Figs. \ref{fig1}  and  \ref{fig2}  show the radio contour map overlaid on the optical images.   The 1.4 GHz radio flux densities is adjusted to the NVSS flux densities considering the errors in both observations.  With a faint source detected at 610 MHz for this source, we obtain a positive spectral index of $\alpha=+0.93$.
\noindent Positive spectral index is usually observed in GHz Peak spectrum sources (GPS) \citep{Stawarz08} or compact steep spectrum sources (CSS) and there are two possible explanations of the observed spectral index, the synchrotron self-absorption and the free-free absorption.

	At low frequency, the continuum spectrum of the radio sources turn from steep or flat spectrum to a spectrum with positive index. This behaviour can be explained with the absorption of synchrotron radio emission by synchrotron electrons. In low frequency where the brightness temperature, ${T}_{b}={\frac{{\lambda}^{2}}{2k}}{\frac{{S}_{\nu}}{\Omega}}$, of radio source is high enough to be equal to its thermal temperature, the emissions is absorbed by free electrons. This occurs in the centimetre and the millimetre wavelengths of nuclei of active galaxies and the quasars. Theoretically the index can reach to +5/2 in frequencies lower than turnover frequency but this is not exactly achieved in observations because of other processes of absorptions and emissions.

In the case of synchrotron self absorption, the continuum spectrum changes from simple form of $f{\propto}{\nu}^{\alpha}$ into 
\begin{eqnarray}
S_{\nu}=S_{m}{({\frac{\nu}{\nu_{m}})}^{\frac{5}{2}}}{\frac{1-e^{-{\tau}_{0}}{({\frac{\nu}{\nu_{m}})}^{{\alpha}-{\frac{5}{2}}}}}{1-e^{-{\tau}_{0}}}}
\end{eqnarray}
which $S_{m}$ and ${\nu}_{m}$ are flux density and frequency of  maximum point of the spectrum  (Megn 2008; Scott \& Readhead 1977) and ${\tau}_{0}$ is determined from the following relation 
\begin{eqnarray}
e^{-{\tau}_{0}}=1+{\frac{-2{\alpha}+5}{5}}{\tau}_{0}
\end{eqnarray}

In the case that synchrotron self absorption is the origin of observed spectral index, we assumed this source followed typical spectrum that other fossils in our sample had, with 
the spectral index approximately equal to $-1$ and fitted  equation (1) to the observed flux densities and found the turnover point parameters, $S_{m} $ = 15.0 mJy and ${\nu}_{m}$ = 1.1 GHz, for the J1331 spectrum. 
The results remain unchanged if we assume ${\alpha}=-$1.2. Thus we are observing near the peak frequency of the spectrum.
Although the peak frequency is close to 1.4 GHz, which is our observed frequency, the source 
size at 1.4 GHz is not necessarily a true maximum angular size for the emitting region, because of the resolution limitations. 
We deduced an estimation for the angular size of this region from the following relation \citep{ Scott77, Kellermann81}
 \begin{eqnarray} 
{\nu}_{m} {\sim} 8 {B}^{1/5}{S}_{m}^{2/5}{\theta}^{-4/5}{(1+z)}^{1/5}GHz
\end{eqnarray}
  Assuming a typical magnetic field,  $B{\sim}{10}^{-4}$ G,  the corresponding angular size of the emitting region 
will be 0.1 mas or 0.2 pc in physical scale. Thus all the emission is originated from the core of the AGN. We find a brightness 
temperature of an order of ${10}^{12}$ K that is consistent with the kinetic temperature of relativistic electrons. While the J1331 is surprisingly luminous at 1.4 GHz, within our sample of fossil galaxy groups, no radio lobe is detected around the central source, a possible explanation would be, that the AGN activity is very recent. Thus, hot electrons of exploding lobes, self absorb the radio emission at low frequencies, which still produce strong emission at 1.4 GHz.

The second alternative to explain the inverse spectrum is free-free absorption. This is due to the absorption of low frequency radio emissions by free electrons of ionised regions in the galaxy medium. In this scenario, the ratio of the observed flux densities at two different frequencies are given by \citep{Levinson95}
\begin{eqnarray}
{\frac{S_{{\nu}_{1}}^{ob}}{S_{{\nu}_{2}}^{ob}}}\approx({\frac{{\nu}_{1}}{{\nu}_{2}}})^{\alpha}exp({-{\Delta}{{\tau}_{ff}}})
\end{eqnarray}
The optical depth ${\tau}_{ff}$ is descripted by the following equation \citep{Tingay01}. 
\begin{eqnarray}
{\tau}_{ff}=5.5 {l}_{pc} {{T}_{4}^{-3/2}} Z {n}_{e4} {n}_{i4} {\nu}_{9}^{-2} {g}_{ff}({\nu},T,Z)
\end{eqnarray}
${l}_{pc}$ is the path length in free-free region in parsec, T is the temperature in units of ${10}^{4}$ K,
$Z$ is equal to one for pure hydrogen plasma, ${n}_{e4} ,{n}_{i4}$ are the electrons and ions number densities
in units of  ${10}^{4} ${cm}$^{-3}$ that we assume to be equal,
${\nu}$ in GHz, and ${g}_{ff}({\nu}_{9},T_{4}) {\sim} 5.99{T}_{4}^{0.15}{\nu}_{9}^{0.1}$ \citep{Brown87}.
Using equations 4 and 5, we derived a constrain in density and size of the absorbing region:
\begin{eqnarray}
{l}_{pc}=0.02  {n}_{e4}^{-2} {{T}_{4}^{-1.35}}
\end{eqnarray}

  There are several possible sources for free-free absorption in this galaxy group.
Absorption of radio emissions by free electrons is significant when we observe the central AGN 
through the disk \citep{Jones00}. Thus one explanation is centred around the line of sight argument toward this galaxy group. Furthermore, cooling flow toward the core of galaxy group provides a reservoir of free electrons that 
could absorb the radio emissions. Following equation (6), we can estimate the length scale (${l}_{pc}$). 
Based on the temperature and electron density of the X-ray gas (${\sim}$ 0.001 {cm}$^{-3}$), the
absorbing length scale would be much bigger than galaxy group dimensions and therefore it is unlikely that these electrons can be seen as a viable source for free-free absorption.

In addition to this positive spectral index, the central galaxy at the X-ray peak, has narrow {H}$_{\alpha}$ and S[II]
emission lines \citep{kpj07}. These emission lines have not been detected for other galaxies in our sample. Thus another possibility is the existence of star forming regions (HII regions) which are responsible for the observed spectral lines and 
 free-free absorption of radio emissions due to their free electrons.
Assuming a typical temperature (${\sim}{10}^{4}$ K ) and density (${\sim}{10}^{3}${cm}$^{-3}$) 
for the HII region, equation (6) gives an estimate (${\sim}$ 2 pc ) for the size of the absorbing region.

\subsection{RX J1552.2+2013}

This galaxy group is the second distant object in our sample and the most luminous object in $K$-band ($M_{k}$= $-$25.7). In addition, \cite{kpj07} described further the properties of this galaxy group in X-ray and optical. \cite{Mendes06} studied the luminosity function of the group and confirmed 36 group members.
At 610 MHz and 1.4 GHz frequencies this group does not show any radio emission at noise level of 80 $\mu$Jy {beam}$^{-1}$. Upper limit for the radio luminosity of $L_{radio}<9.0{\times}10^{21}$ W {Hz}$^{-1}$  for this group is estimated by assuming a 3$\sigma$ cut of the GMRT observations. Such luminosities could indicate a quit AGN at the centre of this group.

\subsection{RX J1416.4+2315}

\begin{figure}
\center
\epsfig{file=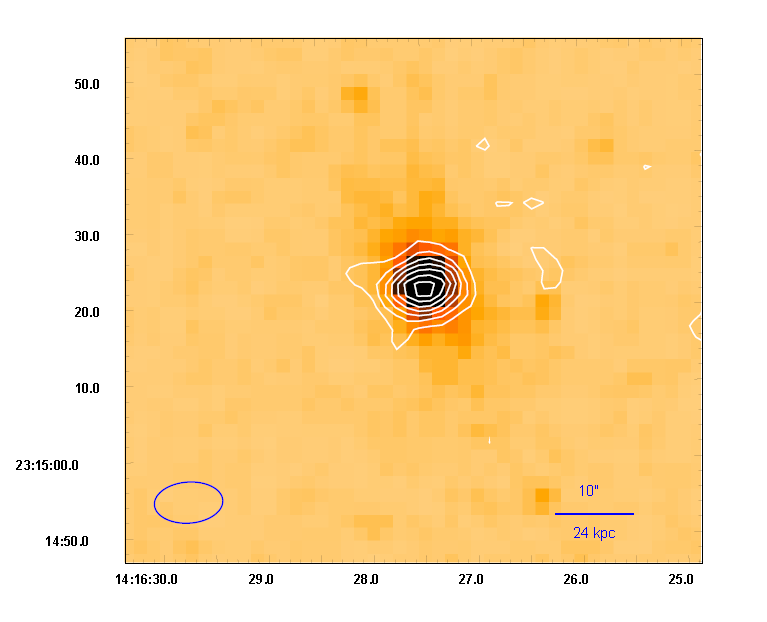 ,width=3.2in,height=2.6in}
\caption{The 1.4 GHz radio map of J1416, overlaid on DSS image of the central galaxy. 
Contour levels of ${\sigma}$ = 0.12 mJy ${\times}$
  3 , 5 , 7 , 9 , 11 , 13 , 17 , 21 , 25 , 29 , 33 are shown. }
\label{fig3}
\end{figure}

\begin{figure}
\center
\epsfig{file=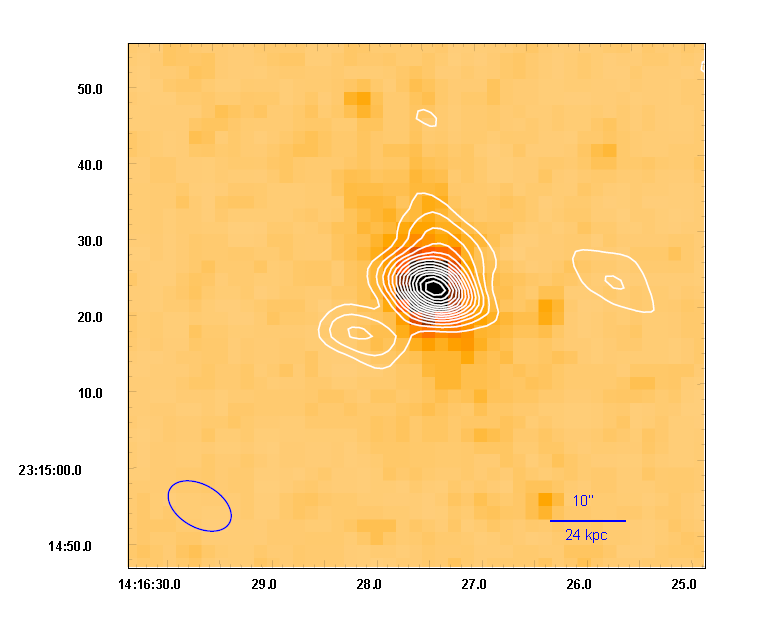,width=3.2in,height=2.6in}
\caption{The 610 MHz radio map of J1416, overlaid on DSS image of the central galaxy. 
Contour levels of ${\sigma}$ = 0.12 mJy ${\times}$
  3 , 5 , 7 , 9 , 11 , 13 , 17 , 21 , 25 , 29 , 33 are shown.}
\label{fig4}
\end{figure}

This object is the most massive and the richest system in our sample and dubbed as a fossil galaxy cluster instead of a group \citep{kmpj06}. The radio analysis is reported extensively in \cite{miraghaei12} and thus only the results are discussed in this paper. Its recent AGN activity is confirmed from radio observation at 610 MHz by detecting a radio lobe of about ${\sim}$30 kpc away from the central source \citep{Jetha09}. Figs. \ref{fig3} and \ref{fig4} show the radio contour map overlaid on the optical images.  The size of the radio lobe is the same as the synthesised beam as shown in Fig. \ref{fig4}. This lobe could heat the IGM by mechanical heating and this possibility will be discussed in section \ref{sec:Heating}.

\subsection{NGC 6482}

\begin{figure}
\center
\epsfig{file=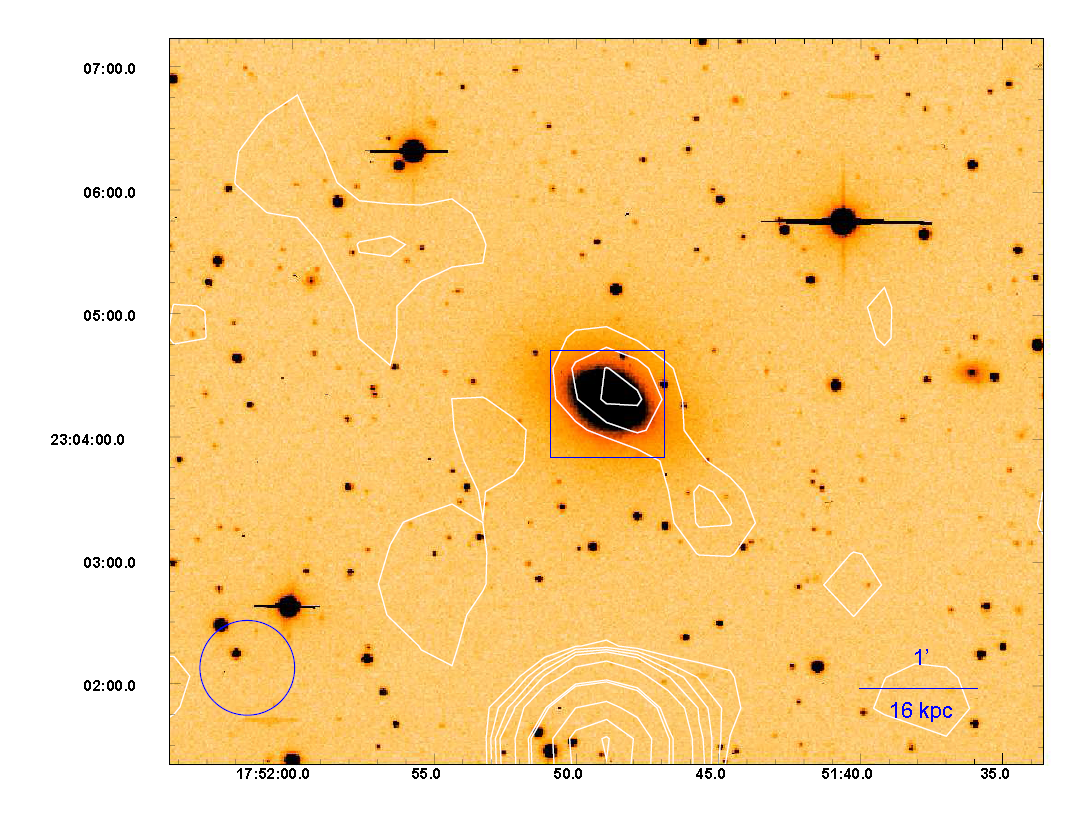,width=3.2in,height=2.5in}
\caption{The 1.4 GHz radio map of NGC 6482, overlaid on $R$-band image of the central galaxy. 
VLA contour levels of ${\sigma}$ = 0.5 mJy ${\times}$
 1 , 2 , 3 , 4 , 8 , 16 , 20 , 40 , 60  are shown. There is a small
peak above  3${\sigma}$ in the galaxy position. High resolution image of the group inside the white box
is illustrated in Fig. \ref{fig6} .}
\label{fig5}
\end{figure}

\begin{figure}
\center
\epsfig{file=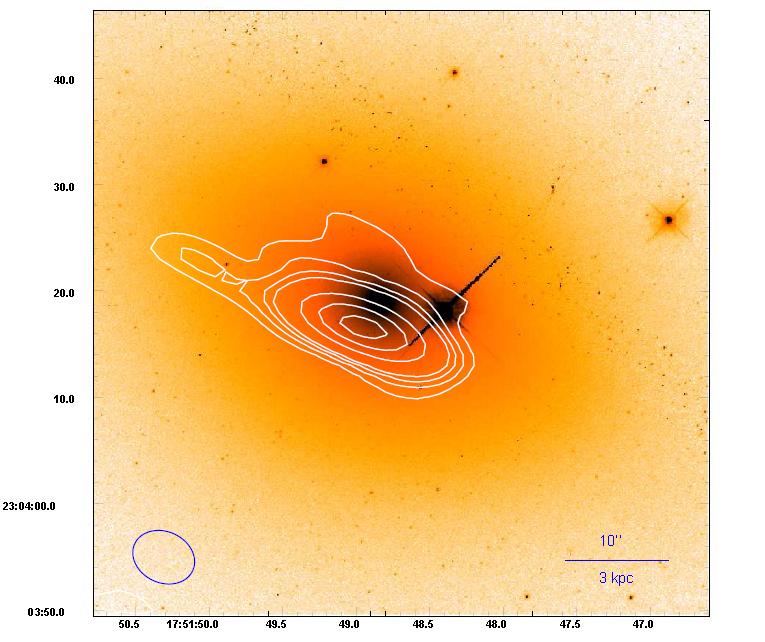,width=3.7in,height=3.2in}
\caption{The 610 MHz radio map of NGC 6482, overlaid on high resolution HST image of the central galaxy. 
Contour levels of ${\sigma}$ = 0.18 mJy ${\times}$
 2 , 3 , 4 , 5 , 10 , 20 , 30 , 40  are shown.}
\label{fig6}
\end{figure}

This galaxy group is the nearest fossil galaxy group found to date, with bolometric X-ray luminosity of
 $1.3{\times}{10}^{43}$ erg s$^{-1}$ and $R$-band magnitude of $-$21.36, it is considered as the poor fossil galaxy group of our sample. The X-ray analysis of this system has been reported by \cite{kjp04}.  The temperature profile of this group shows no sign of a decrease at the centre albeit the cooling time dropping below the Hubble time within a 36 kpc radius . At 610 MHz (Fig. \ref{fig6}), a faint radio source as luminous as 5.38 mJy and an angular scale size of ${\sim}$10 arcsec (${\sim}$2 kpc) has been detected. The radio emission originated from the central region (as determined by the optical image ) of the galaxy with corresponding luminosity of  $L_{radio}=0.19{\times}10^{22}$ W {Hz}$^{-1}$ . The 1.4 GHz map, adopted from the VLA archive shows flux densities at 3${\sigma}$ level at the position of NGC 6482 (Fig. \ref{fig5}). Based on this coincidence we assumed, that this radio emission seems to be a real detection and estimate a spectral index of $\alpha_{610}^{1420}$=$-$1.05.

\subsection{ESO 3060170}
\label{sec:Result5}
The ESO 3060170, as a massive fossil galaxy group, has been extensively described in \cite{Sun04}. One of their findings are, that a heating mechanism is needed in this group to balance the X-ray energy loss in the small cool core ($\sim$ 10 kpc) at the centre of this group, compared to its overall $\sim$ 50 kpc cooling radius.
The radio maps at 1.4 GHz and  610 MHz are presented in Figs. \ref{fig8}  and \ref{fig9}, respectively.  
In 610 MHz map, there is a core radio emission with a flux density of 15.06 mJy, a small radio extension
with an integrated flux density of 2.6 mJy toward the East.
The peak flux density of radio lobe is 2.23 mJy {beam}$^{-1}$, at a 7$\sigma$ ($\sigma$ = 0.3 mJy {beam}$^{-1}$) level
of the map around the lobe, thus it is a radio bubble for the central galaxy.
The source is the same size as the synthesized beam. In 1.4 GHz map, the same extension is detected with a peak flux density of 0.66 mJy {beam}$^{-1}$ for the radio lobe, more than  3$\sigma$ of the map with an integrated flux density of 1.34 mJy.
Both the radio lobe and the central source have a similar spectral index (${\sim} -$0.9). 
There is also small extension to the West of the central region, more visible in 1.4 GHz map
with the peak flux density of 2$\sigma$, so with a 95 percent probability is regarded as a second lobe originated from the AGN but the size of this extension is smaller than the synthesized beam, thus we are unsure if this is a real detection.  
The Eastern radio lobe is separated about 6 arcsec (${\sim}$ 4 kpc) from the centre so it is entirely inside the cool core of ESO 3060170. In section \ref{sec:Heating1}, we estimate the energy budget of the radio lobe for heating up the IGM.

\begin{figure}
\center
\epsfig{file=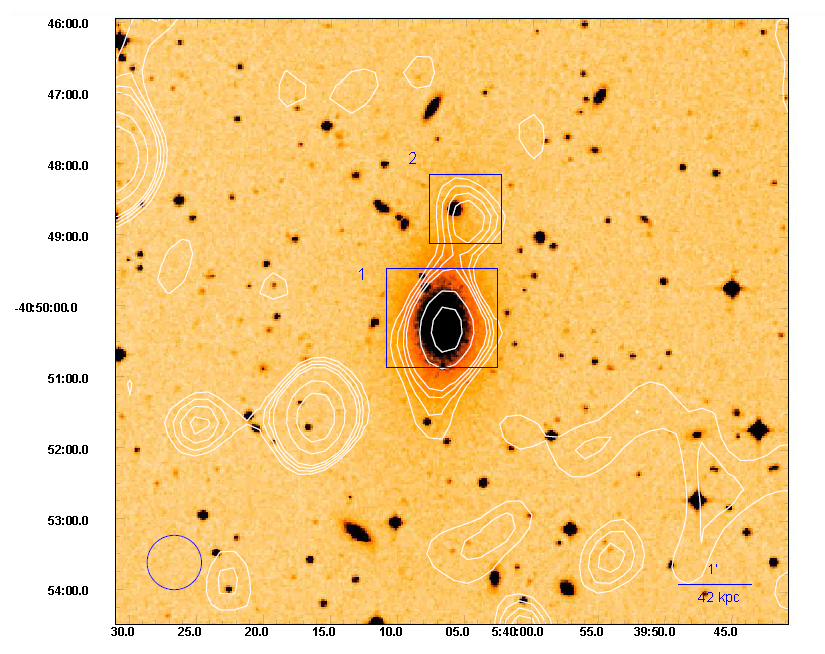,width=3.5in,height=2.8in}
\caption{The 843 MHz SUMSS radio map of ESO 3060170, overlaid on DSS image of the central galaxy. 
The bubble like extension on top 
of the galaxy (box 2) is radio emission from its neighbour that misunderstood
as a radio lobe for ESO 3060170 in previous works.
High resolution map of 610 MHz in Fig. \ref{fig10}  discarded this idea. Figs. \ref{fig8}  and \ref{fig9}  show the central part of
group inside box 1 and ESO 3060170 radio lobes.  }
\label{fig7}
\end{figure}

\begin{figure}
\center
\epsfig{file=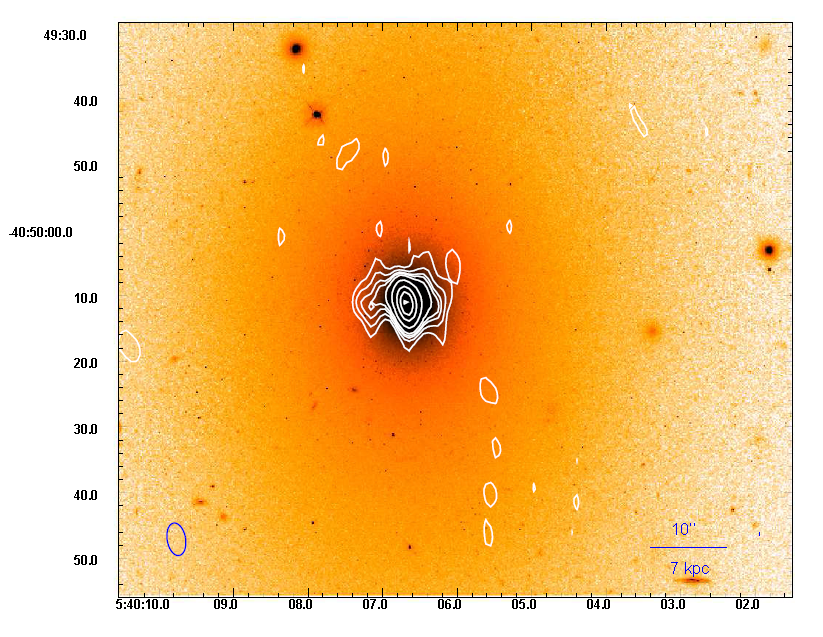,width=3.4in,height=2.9in}
\caption{The 1.4 GHz radio map of ESO 3060170, overlaid on high resolution HST image of the central galaxy. 
Contour levels of ${\sigma}$ = 0.15 mJy ${\times}$
 1 , 2 , 3 , 4 , 5 , 10 , 20 , 30  are shown. A radio lobe at the east 
of central part and a small extension on the west show its recent AGN activity.}
\label{fig8}
\end{figure}

\begin{figure}
\center
\epsfig{file=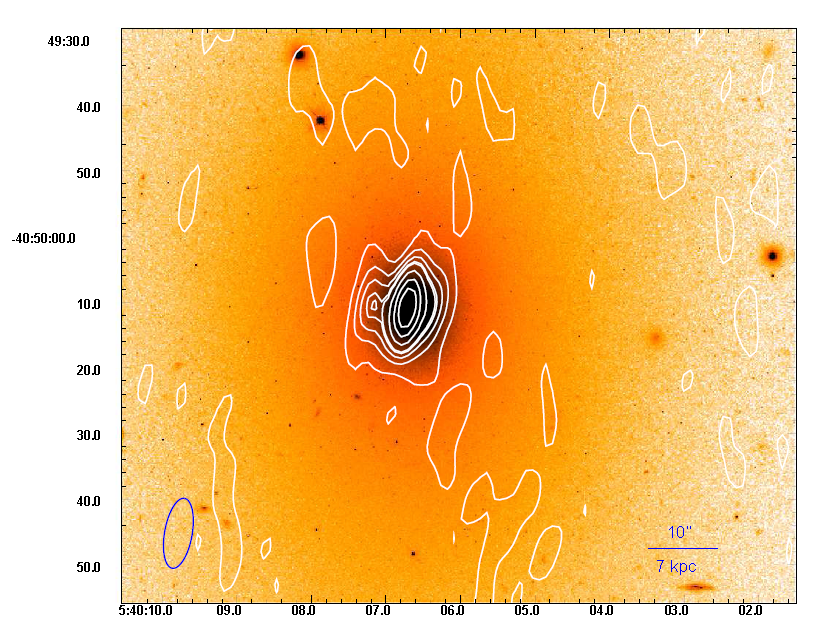,width=3.4in,height=2.9in}
\caption{The 610 MHz radio map of ESO 3060170, overlaid on high resolution HST image of the central galaxy.
Contour levels of ${\sigma}$ = 0.3 mJy ${\times}$
 1  , 3 ,  5  , 7 , 8 , 12 , 18 , 25  are shown. A radio lobe at the east 
of central part similar to 1.4 GHz map is detected.}
\label{fig9}
\end{figure}

\begin{figure}
\center
\epsfig{file=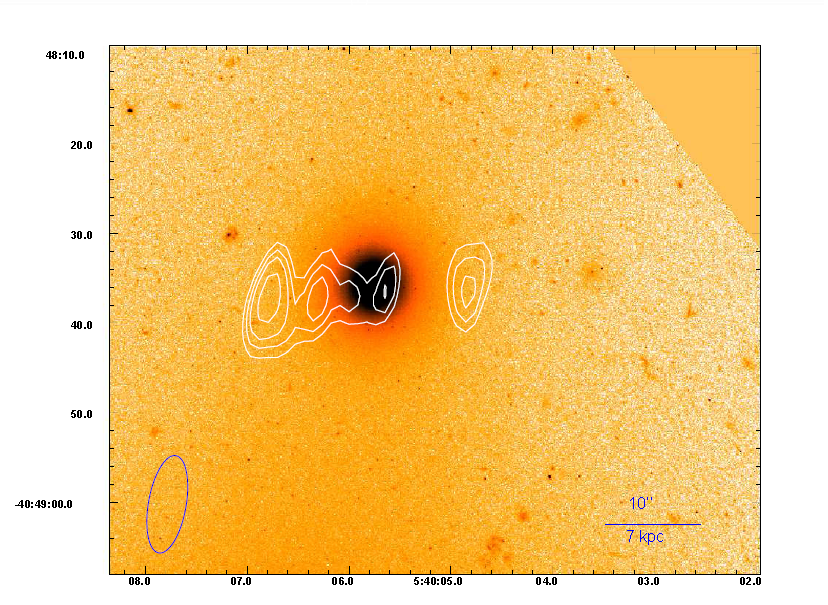,width=3.4in,height=2.9in}
\caption{2MASX J05400579-4048354, the galaxy at the north of ESO 3060170
that its radio emission is mistaken as ESO 3060170 radio lobe in low 
resolution map (see Fig. \ref{fig7} ).}
\label{fig10}
\end{figure}

There is a finger shape extension in the X-ray map of ESO 3060170, pointing to the north west  
with the scale of 30 kpc (see  \cite{Sun04} ) which is possibly related to the AGN outbursts, 
however, there is no sign of an AGN activity in this direction in the radio maps. 
The radio map obtained from the SUMSS archive at 843 MHz (Fig. \ref{fig7} ) shows a bubble like source (box 2) in the same direction of the X-ray extension which is identified as a radio bubble related to the central galaxy in \cite{Sun04}.
As shown in Fig. \ref{fig7}, there is a galaxy (2MASX J05400579-4048354) at the north of the central galaxy where the claimed bubble
is present (box 2). Fig. \ref{fig10} shows the zoomed in radio map of the box 2 region at 610 MHz. It is clear that the origin of this 
radio emission is the galaxy itself. As the map shows, this galaxy (2MASX J05400579-4048354) may have had several cycles of outbursts.

In Table 5, we listed the radio flux densities of ESO 3060170 from our observations and publicly available surveys. 
The PMN resolution is 5 arcminutes, so the radio flux density is the integrated flux densities of ESO 3060170
and other nearby sources within 5 arcminutes. We  exclude this measurement.
The SUMSS and the GMRT data show that ESO 3060170 have almost flat radio spectrum below 843MHz (SUMSS rms noise ${\sim}$ 1 mJy {beam}$^{-1}$).
 The spectral index between 834 MHz and 1.4 GHz is ${\alpha}=-1.7$.

\begin{table}
\begin{center}
\caption{ESO 3060170 radio spectrum 
\label{table5}
}
\begin{tabular}{lll}
\hline
$\nu$ & f& Survey/Telescope \\
 $[$GHz$]$ & [mJy] &  \\ 
\hline 

4.58&48 &PMN\\
0.843&19 &SUMSS\\
1.4&9.4&GMRT\\
0.61&17.66&GMRT\\

\hline  
\end{tabular}
\end{center}
\end{table}

\section{IGM Heating}
\label{sec:Heating}

In this section, we examine mechanical heating, conduction, shock and vortex heating to investigate if these heating processes will provide sufficient energy to balance the cooling rate in our sample.

\subsection{Mechanical Heating}
\label{sec:Heating1}
There are a number of galaxy clusters in which X-ray cavities, known as bubbles, are formed around the brightest cluster galaxy which hosts a massive black hole \citep{Tremblay12, Randall11, Gitti10, Birzan04, McNamara05}. The bubbles are filled with the injected relativistic electrons originated from the AGN.  They expand until they reach pressure equilibrium with the surrounding environment and start their journey throughout the IGM up to hundreds of kpc away from the centre. This picture is supported by a number of studies focused on the efficiency of this type of AGN feedback by using simulations \citep{Sijacki06, Okamoto08}. Thus PdV work of AGN outflows on the IGM is considered as a possible source of heating and a thorough estimation of the PdV work taking place in our sample is needed.

In general, the total energy of a bubble is approximated by bubble enthalpy which is the internal energy and the PV work \citep{Churazov02}.
\begin{eqnarray}
H=E+PV
\end{eqnarray} 
In this study, we adopted enthalpy changes of radio bubble as an estimate for the energy release when a bubble moves from
the core of galaxy group or cluster to its present location.

 In the galaxy groups ESO 3060170 and J1416, the existence of radio lobes near the central galaxies are discussed in section \ref{sec:Result}. The energetic discussions for J1416 is reported in a separate study \citep{miraghaei12}. However using a similar method, we evaluated the mechanical work done by the radio lobe of ESO 3060170 and compared this energy with the energy loss from X-ray emission. 
\begin{table*}
\begin{center}
\caption{Energy injection in the core of ESO 3060170 
\label{table6}
}
\begin{tabular}{lllllllllr}
\hline
overpressure & $t_{r}$ & $t_{b}$ & $t_{c}$  & ${\Delta}H$ & $P_{r}$ & $P_{b}$ &$P_{c}$& \\
    factor    & [{10}$^{6}$yr]  & [{10}$^{6}$yr]  & [{10}$^{6}$yr]    & [{10}$^{56}$erg]   &[{10}$^{42}$erg s$^{-1}$]&[{10}$^{42}$erg s$^{-1}$]&[{10}$^{42}$erg s$^{-1}$]&&\\ 
\hline 
2&6.3&3.1&7.2&1.24&0.62&1.29&0.54&&\\
10&6.3&3.1&7.2&5.11&2.57&5.31&2.24&&\\
\hline  
\end{tabular}
\end{center}
\end{table*}
\begin{table*}
\begin{center}
\caption{Heating vs Cooling 
\label{table7}
}
\begin{tabular}{lll}
\hline
$ $ & ESO 3060170& J1416 \\

\hline 
$R_{c}$&50 kpc &130 kpc\\
$R_{cc}$&10 kpc&50 kpc\\
$Lx<R_{c}$& 5.9${\times}${10}$^{43}$erg s$^{-1}$&2.5${\times}${10}$^{43}$erg s$^{-1}$\\
$Lx<R_{cc}$ &7.7${\times}${10}$^{41}$erg s$^{-1}$&2.8${\times}${10}$^{42}$erg s$^{-1}$\\
$d$ &4 kpc& 28 kpc\\
$L_{mech}$ & 0.5-5.3${\times}${10}$^{42}$erg s$^{-1}$&2.3-12.5${\times}${10}$^{42}$erg s$^{-1}$\\
\hline  
\end{tabular}
\end{center}
\end{table*}
We define two cooling radius within which the X-ray loss are determined. First the radius that
cooling time falls below the Hubble time, $R_{c}$, and second the radius at which the temperature
drops indicating the presence of a cool core, $R_{cc}$ ($R_{cc}<R_{c}$).  For ESO 3060170, there is a 
clear temperature drop below 10 kpc while for J1416 there is just a slight decrease in the temperature 
profile below 50 kpc.  The bolometric X-ray luminosity within both radii are determined and reported in Table 7. 
To calculate heating rate of radio bubbles, we adapted three different time scales introduced by \cite{Birzan04},
all listed in Table 6. We calculate the mechanical heating assuming the over pressure factors  (${\frac{p_{0}}{p}}$)  
of 2 and 10 for the expanding bubble \citep{Jetha08}. 
In this way, the enthalpy change is \citep{miraghaei12}:
\begin{eqnarray}
{\Delta}H={\frac{\gamma}{{\gamma}-1}}PV[{({\frac{p_{0}}{p}}})^{1-1/{\gamma}}-1]
\end{eqnarray}
${\gamma}$ is the adiabatic index , P and V are IGM pressure and the bubble volume, respectively. 
For ESO 3060170, mechanical power changes from 0.5  to   5.3${\times}{10}^{42}$ erg s$^{-1}$ 
applying different time scales and over pressure factors. Although this is quite sufficient to stop cooling within the $R_{cc}$ with X-ray loss of  ${\sim}{10}^{41}$ erg s$^{-1}$, it is not comparable to the X-ray loss within $R_{c}$, with ${L}_{x}{\sim}{10}^{43}$ erg s$^{-1}$. As it is shown in Table 7, this is also seen in J1416.  Thus  a more energetic ( e.g. for ESO 3060170, $>{10}^{57}$ erg ) or rapid (${\sim}{10}^{5}$ yr) cycle of AGN activity 
is needed.  The latter  is not typical for AGNs while the former is probable.

\begin{figure}
\center
\epsfig{file=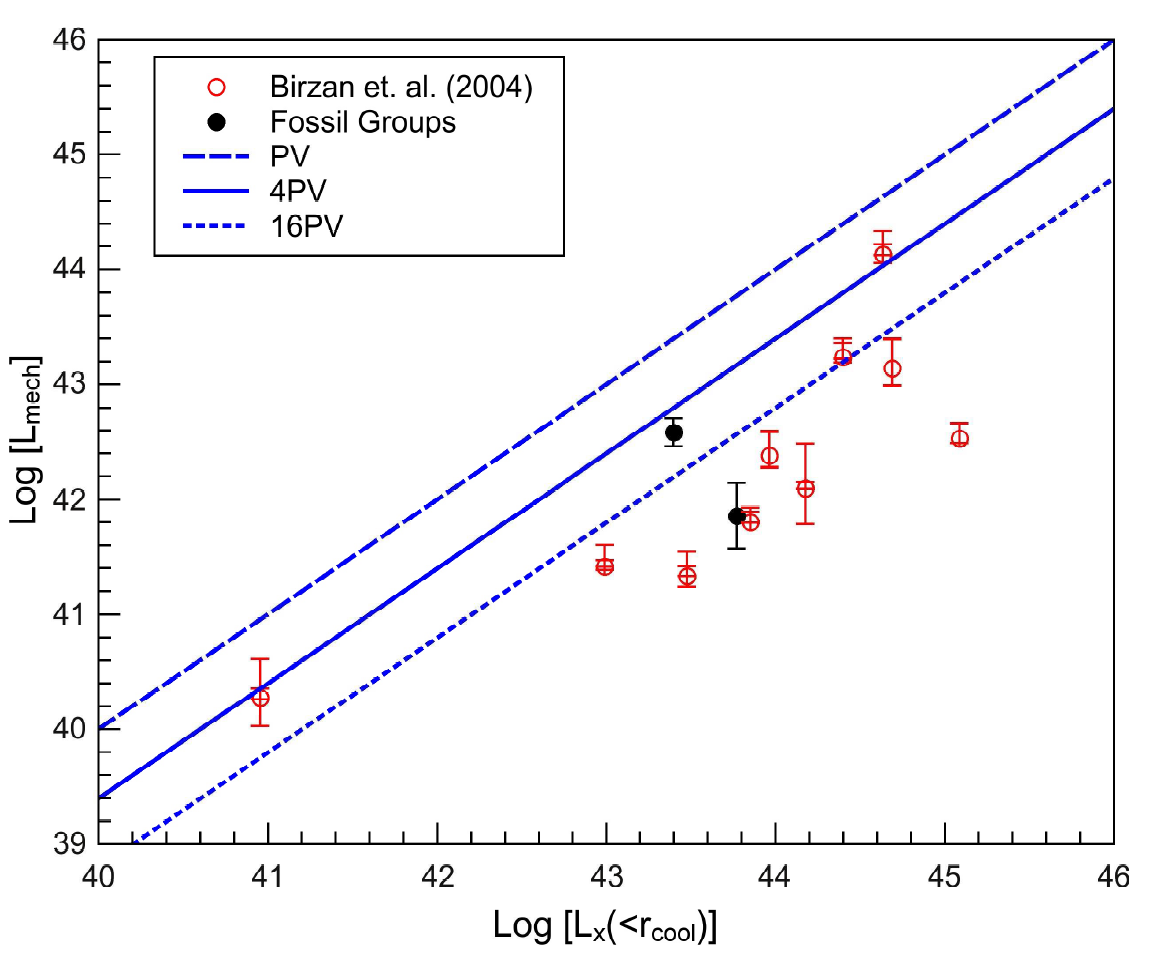,width=3.2in,height=3.2in}
\caption{Mechanical heating vs. cooling rate within cooling radius for \citet{Birzan04} normal groups and our fossil groups.
PV work are estimated only for the radio filled cavities in both samples.}
\label{fig15}
\end{figure}

\cite{Birzan04} show, for a sample of galaxy clusters, that mechanical heating
is efficient if they assumed 1 PV to 16 PV energy per cavity. In our sample 6 PV and 34 PV 
energy per cavity are required to stop the cooling process in J1416 and ESO 3060170, respectively. This is significantly higher than the total enthalpy of a cavity or a radio lobe ($H = 4PV$). So mechanical heating does not seem to be a major contributor to the IGM heating for these objects, however AGN can still be an important sources of heating via other mechanisms (see e.g. the following sections). Fig.~\ref{fig15} shows the $L_{Mech}$-$L_{x}$ plot for the \cite{Birzan04} sample and our fossil objects. Since we detected bubbles at radio band in our sample, we adopted just filled cavities of \cite{Birzan04} sample to make the analogy reliable. The horizontal axis is X-ray luminosities within cooling radius. Our fossil groups are located in poor clusters criteria and fall below 4PV line.  Although the sample size is very small, there is no indication that the mechanical power differs between fossil and non-fossil systems. We further investigate AGN power in fossil and non-fossil systems via their radio luminosities and other clusters and groups characteristics (section \ref{sec:stat}). 
One of the noticeable results in this study is that the two most X-ray luminous and massive objects in our sample just possess AGN radio bubbles. This is consistent with apparent correlation between L$_{Mech}$ and L$_{x}$ shown in Fig. \ref{fig15}. However the plot suffers from weak selection biases \citep{Birzan04}. In section \ref{sec:stat}, we also try to investigate this issue using various samples of galaxy groups and clusters.

In estimating the mechanical heating of radio bubble, we assumed it is in
pressure balance with the surrounding gas, with such assumption one can deduce
magnetic field of the radio lobe \citep{Birzan08}. Thermal pressure out of the bubble should be equal
to particle pressure plus magnetic field pressure inside the lobe. Using this method
we deduced magnetic field of 63 ${\mu}$G for ESO 3060170 comparing to equipartition magnetic field (0.8 ${\mu}$G).

\subsection{Shock and Vortex Heating}
\label{sec:Heating2}
In addition to mechanical heating, AGNs heat the inter galactic gas via 
weak shocks induced by expanding bubbles or large scale shocks \citep{Fabian06, Nulsen05, Forman07}, sound waves \citep{Fabian03} 
and interacting of shocks and bubbles \citep{Friedman12}. Shock energy per volume is approximated by pressure jump 
across the shock front. Therefore we have
\begin{eqnarray}
{\Delta}P{\sim}{\frac{E}{V}}
\end{eqnarray}
Pressure ratio at shocked (${{P}_{2}}$) and unshocked (${{P}_{1}}$) regions are given by
\begin{eqnarray}
{\frac{{P}_{2}}{{P}_{1}}}={\frac{2{\gamma}{M}^{2}+(1-{\gamma})}{({\gamma}+1)}}
\end{eqnarray}
Here ${\gamma}$ is adiabatic index and M is Mach number estimated
from temperature increase using 
\begin{eqnarray}
{\frac{{T}_{2}}{{T}_{1}}}={\frac{[2{\gamma}{M}^{2}+(1-{\gamma})][{\gamma}-1+{\frac{2}{{M}^{2}}}]}{{({\gamma}+1)}^{2}}}
\end{eqnarray}
${{T}_{2}}$ and ${{T}_{1}}$ are the temperatures at shocked and unshocked regions respectively.
The finger shape region in X-ray map of ESO 3060170 that we discussed in section \ref{sec:Result5} contains hotter gas 
compared to its surroundings \citep{Sun04}. We assumed this X-ray enhancement is the consequence of AGN shocks
and estimated Mach number of $\sim$1.2 for that. Concerning its volume, this corresponds to shock heating
of $\sim$1$\times {10}^{42}$ erg s$^{-1}$.

\begin{figure}
\center
\epsfig{file=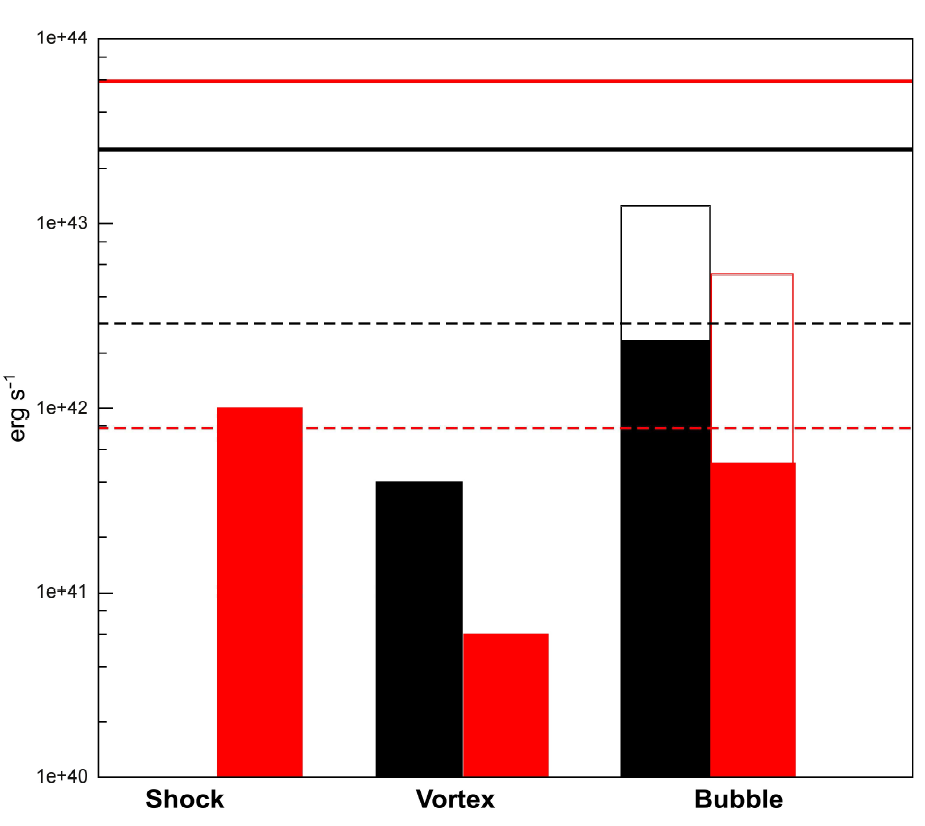,width=3.4in,height=2.9in}
\caption{IGM Heating in fossil groups J1416 (black) and ESO 3060170 (red) via Shock, vortex and bubble heating. Solid and dotted lines correspond to X-ray loss within $R_{c}$ and $R_{cc}$ respectively for each fossil system. The filled and unfilled regions in bubble heating correspond to overpressure factors of 2 and 10 respectively.}
\label{figtest}
\end{figure}

Interaction of radio bubbles of previous AGN cycles with AGN shocks and dissipation 
of their rotational energy in vortices (vortex heating) is another heating
source provided by AGNs \citep{Friedman12}. To examine this idea, we evaluated the energy for our detected radio bubbles
assuming they are confronted AGN weak shocks. The energy in the vortex field is given by
\begin{eqnarray}
E{\approx}{{\rho}} {{{\Delta}{\nu}^{2}}}{V}
\end{eqnarray}
${\rho}$ is the ICM density, V is the bubble volume and  ${\Delta}{\nu}$  is the velocity jump 
calculated by shock mach number.
We evaluated the released energy in time scale 
that the bubble reaches to its current position for both objects. The heating powers estimated for 
ESO 3060170 and J1416 are 6.4${\times}{10}^{40}$ erg {s}$^{-1}$
and 4.2${\times}{10}^{41}$ erg {s}$^{-1}$ respectively.
 Fig. \ref{figtest} illustrates a comparison between three different heating mechanisms supplied by the AGNs for two sample groups. 

\subsection {Thermal Conduction}
 
Conduction is widely discussed in the literature as a uniformly distributed heat supply
 at the centre of cool core galaxy groups and clusters.
This heating mechanism is discussed in the literature in conduction-only models
 \citep{Ruszkowski11, Bregman88, Zakamska03} or hybrid models of
 conduction and AGN heating \citep{Ruszkowski02, Guo08}
to explain heating and cooling balance in the cooling flow
clusters.  The latter would be a good alternative for the models which adapted AGNs the
only source of heating and therefore  suffer from direction dependent heating. 
In section \ref{sec:Heating1} and \ref{sec:Heating2},  we showed how AGNs in our sample could not completely stop the cooling.
Here, we examine thermal conduction as a second source of heating to decide if its power 
is enough, contrary to the AGN power. We used the similar method we applied in the 
previous work for J1416 \citep{miraghaei12}. Here, we work over ESO 3060170 that is classified as a strong 
cool core galaxy group \citep{Hudson10} . 

Temperature profile of ESO 3060170 shows an obvious drop within 10 kpc radius in
the centre of group where the cooling time falls below 1Gyr. Such a cool region needs 
$2-3{\times}{10}^{58}$ erg energy to reach its surrounding temperature.
This energy is one order of magnitude higher than the bubble enthalpy.
So the existence of cool core is not in conflict with the AGN outburst
energy in this case.
The surprising point is that even if this energy was smaller than AGN enthalpy
we still see such temperature gap \citep{OSullivan11}. 
So AGN doesn't play
the main role in this matter. 
Note that strong cool core clusters (SCC) are hosting radio 
load AGNs more frequent than weak (WCC) or non cool core (NCC) clusters \citep{Mittal09} that makes this issue more puzzling.     
In addition to energy budget adequacy to disturb 
cool cores, the distribution  of heating and cooling rates in IGM is also important.
In this way,
the question turns to how heat supply distributes over cluster medium which
observed temperature profile can be described.

To investigate if conduction is efficient as a uniformly distributed
heating mechanism, we deduced
 conduction time 
that is the time consumed to thermalize inside and outside of the core.
For the temperature drop of ${\sim}1$ keV within 10 kpc radius with the electron density of 0.08-0.02 cm$^{-3}$,  conduction time is $t=7{\times}{10}^{7}$ yr.  This time is one order 
of magnitude smaller than the cooling time within $R_{cc}$, so conduction is suppressed 10
times by the magnetic field. The heating power conducted into the centre of galaxy group is
$P_{con}=9{\times}{10}^{42}$ erg {s}$^{-1}$. Comparing this power to the X-ray loss within 10 kpc,   
$L_{x}=8{\times}{10}^{41}$ erg {s}$^{-1}$, it is again confirmed that the conduction is suppressed  one order of magnitude by the 
magnetic field.  Thus observing  such temperature drop proves suppression factor higher than 10
is needed.

\section {Powering the central AGN }  
\label{sec:Accretion}
AGN powerful outbursts are certainly supported by feeding mechanisms that 
are still poorly understood. Accretion of hot gas at the vicinity of the black hole, cold molecular 
gas accretion of the cooling flow, stellar accretion and supper massive black hole spin
are the candidates for feeding central AGNs \citep{McNamara11}.  In this section we investigate different ways of feeding
to find if they are powerful enough to explain the observed AGN outbursts of ESO 3060170 and J1416. 

Bondi Accretion or hot accretion \citep{Bondi52}, a spherically symmetric collapse of hot gas into the
black hole is proportional to the gas temperature, electron density and the black hole mass.
To estimate the central black hole mass we adapted \cite{Marconi03} relations for black hole mass and $K_{s}$ 
luminosity of the galaxy.  For ESO 3060170, $K$-band absolute magnitude of $-$25.98 from 2MASS-XSC results in 
$M_{BH}=1.43{\times}{10}^{9} $M$_{\odot}$, so Bondi accretion rate will be $M_{bondi}=6.8{\times}{10}^{-3}$ M$_{\odot}$ yr$^{-1}$. This rate of mass accreted to the centre of galaxy is equal to Bondi power of 
$3.8{\times}{10}^{43}$ erg {s}$^{-1}$ with the assumption of 0.1 efficiency. This is more
than jet power we evaluated in section \ref{sec:Heating}. 
Thus for ESO 3060170, the Bondi accretion is sufficient to fuel the central AGN. This is not the case for J1416 \citep{miraghaei12}.

\noindent \cite{Pope12} discussed accretion of hot gas into black hole from tenth of 
kilo parsecs away when the gas is gravitationally unstable as a result of radiative cooling.
The corresponding heating rate is proportional to $\sigma^3$, which $\sigma$ is the velocity dispersion.
For our objects, it is 4-5 order of magnitude higher than AGNs powers therefore its efficiency would be too low.
Moreover, mass deposition rates, $\dot{M}_{cool}$, of isobaric cooling in the core of galaxy groups
are $\sim$ 0.2 M$_{\odot}$ {yr}$^{-1}$ and $\sim$ 0.3 M$_{\odot}$ {yr}$^{-1}$ for ESO 3060170 and J1416 respectively.
These rates are related to the power of $\sim {10}^{46}$ erg {s}$^{-1}$ that is again far more than the AGNs powers. 
Cold molecular gas in the BCGs or the gas provided by cooling flow, accreted into the black hole also
produce enough energy (e.g $\sim {10}^{63}$ erg for ESO 3060170 ) to light up AGNs.
But lack of correlation between molecular gas and AGNs jet power  \citep{McNamara11}
make this idea uncertain. In addition to that the way that such cold clumps conducted 
into the super massive black holes is still not clear \citep{Pizzolato10}. 

Besides the accretion mechanisms, spinning black hole with a 
rotating parameter, $a$ ($0<a<1$), and black hole mass of $M$, provides 
the energy \citep{Meier99}: 
\begin{eqnarray}
{E}_{rot}{\sim}1.6{\times}{10}^{62}{M}_{9}{a}^{2} erg
\end{eqnarray}
We evaluated this energy for our objects and compared it to the total enthalpy of 
the radio bubbles. For both objects rotating parameter of ${\sim}0.002$ is consistent
with the observed jet energy. Although a maximally 
rotating black hole , $a = 1$, provide thousands times of this energy. 

\section{A statistical comparison}
\label{sec:stat}
As mentioned before, galaxy merger as a direct heating source could provide huge amount of energy into the IGM. It has also been shown recently that a large offset between the X-ray emission and the location of the brightest cluster galaxy as a proxy for merging galaxy systems is correlated with the lack of strong cool core in the cluster \citep{Sanderson09}. This is understood to be direct consequence of mergers within the group/cluster environment. Galaxy mergers can also fuel the AGN, thus indirectly contribute to the IGM heating. To examine the latter, we compare the radio luminosity of fossil groups as non-merging systems with a general population of galaxy groups and clusters. In addition, we explore how the radio luminosity in fossil groups scales with other group properties in comparison to other  group/cluster samples. To this aim, we use both bolometric and monochromatic radio luminosity together with group overall characteristics and their BGG (BCG) characteristics.
 
\subsection{Radio luminosity in fossil and non-fossil groups}
\label{sec:OldAGN}

\begin{figure}
\center

\epsfig{file=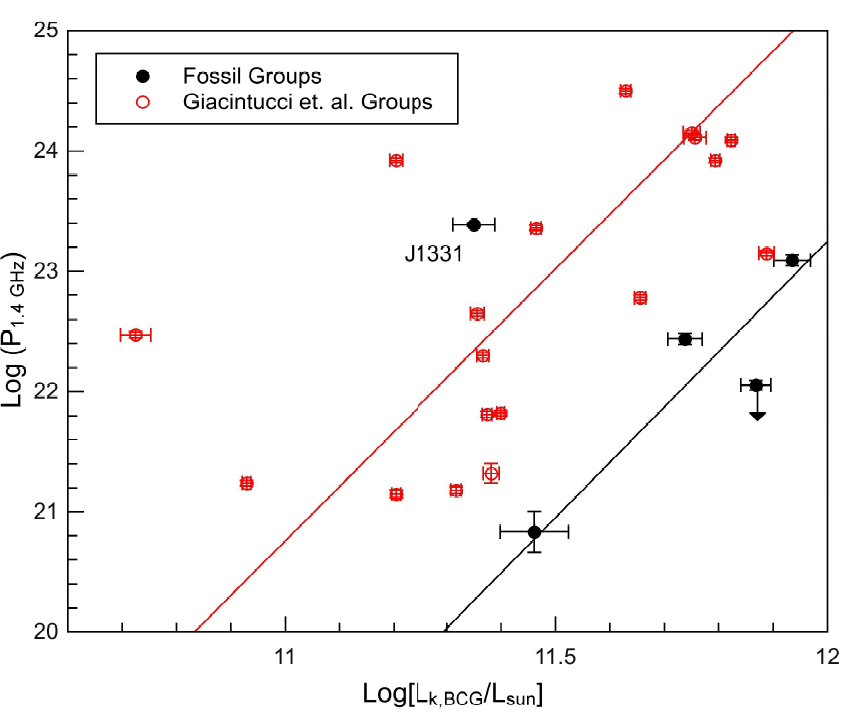,width=3.4in,height=3.2in}
\caption{1.4 GHz luminosity vs $k$-band BGGs luminosity of
fossil galaxy groups (black), and Giacintucci et al. (2011)
groups (red). The red and black lines correspond to fits to the red and black samples.}
\label{fig11}
\end{figure}

\begin{figure}
\center
\epsfig{file=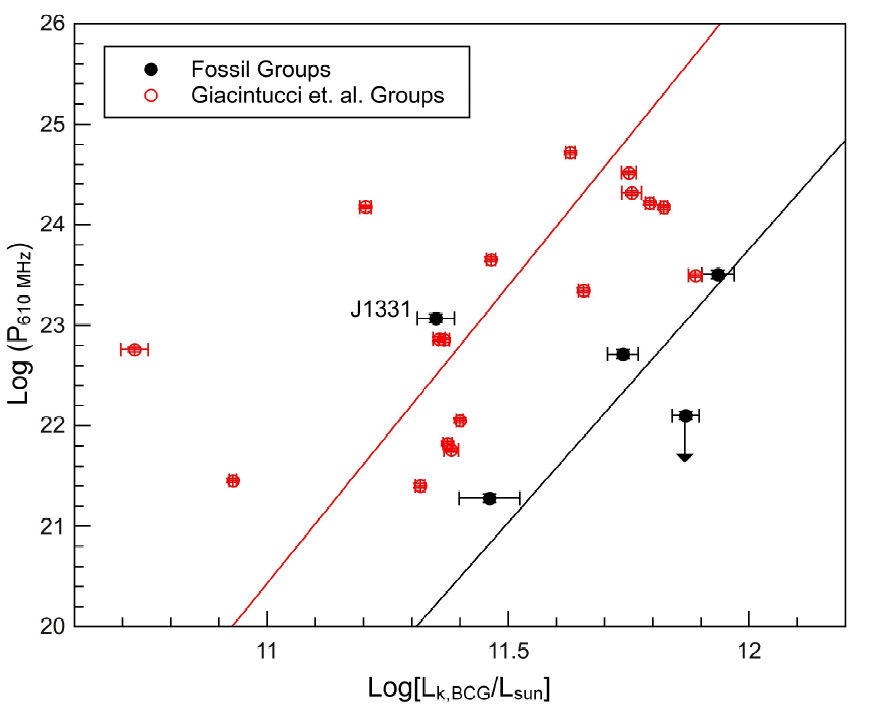,width=3.4in,height=3.2in}
\caption{610 MHz luminosity vs $k$-band BGGs luminosity of
fossil galaxy groups (black) and Giacintucci et al. (2011)
groups (red). The red and black lines correspond to fits to the red and black samples.}
\label{fig12}
\end{figure}

\begin{figure}
\center
\epsfig{file=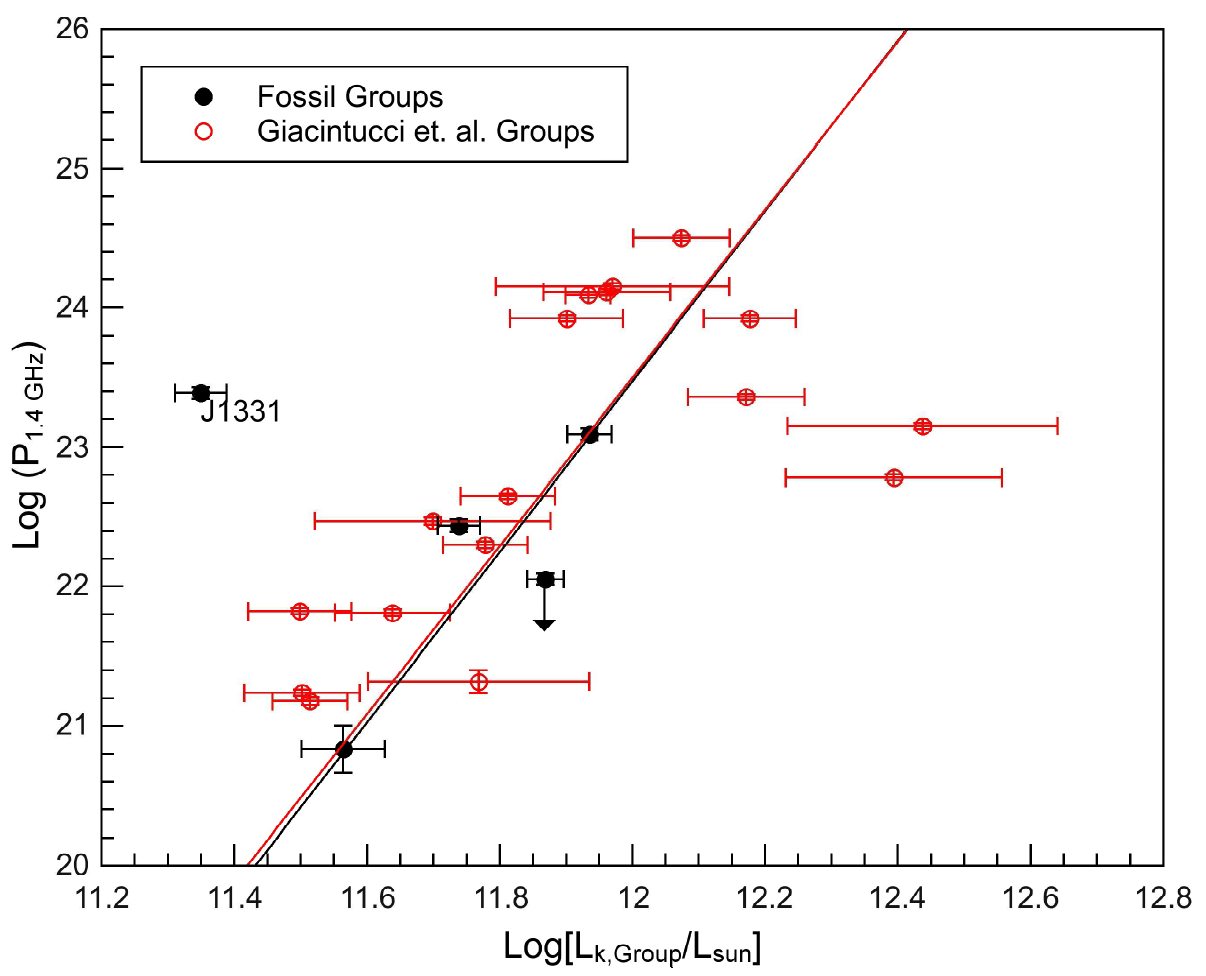,width=3.4in,height=3.2in}
\caption{1.4 GHz luminosity vs $k$-band group luminosity of
fossil galaxy groups (black), and Giacintucci et al. (2011)
groups (red). The red and black lines correspond to fits to the red and black samples.}
\label{fig13}
\end{figure}

\begin{figure}
\center
\epsfig{file=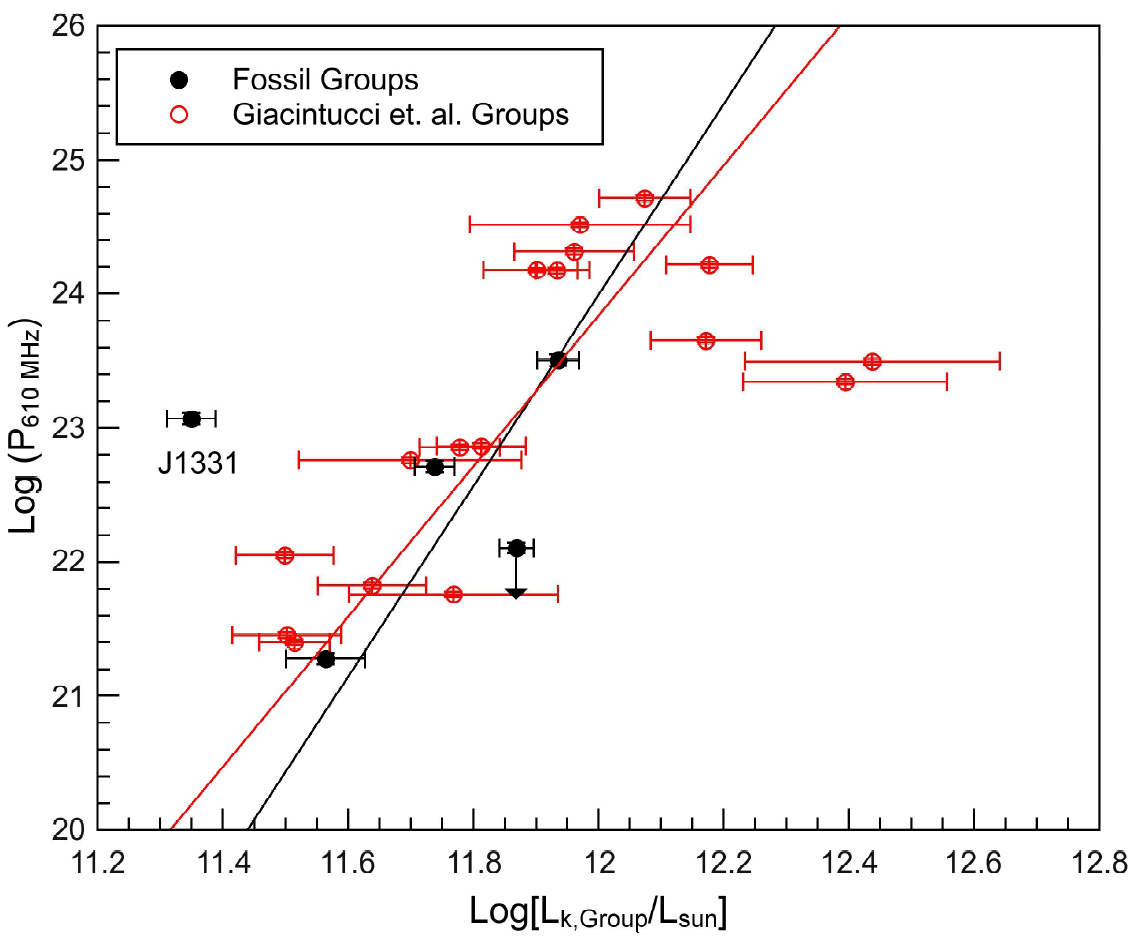,width=3.4in,height=3.2in}
\caption{610 MHz luminosity vs $k$-band group luminosity of
fossil galaxy groups (black) and Giacintucci et al. (2011)
groups (red). The red and black lines correspond to fits to the red and black samples.}
\label{fig14}
\end{figure}

\begin{figure}
\center
\epsfig{file=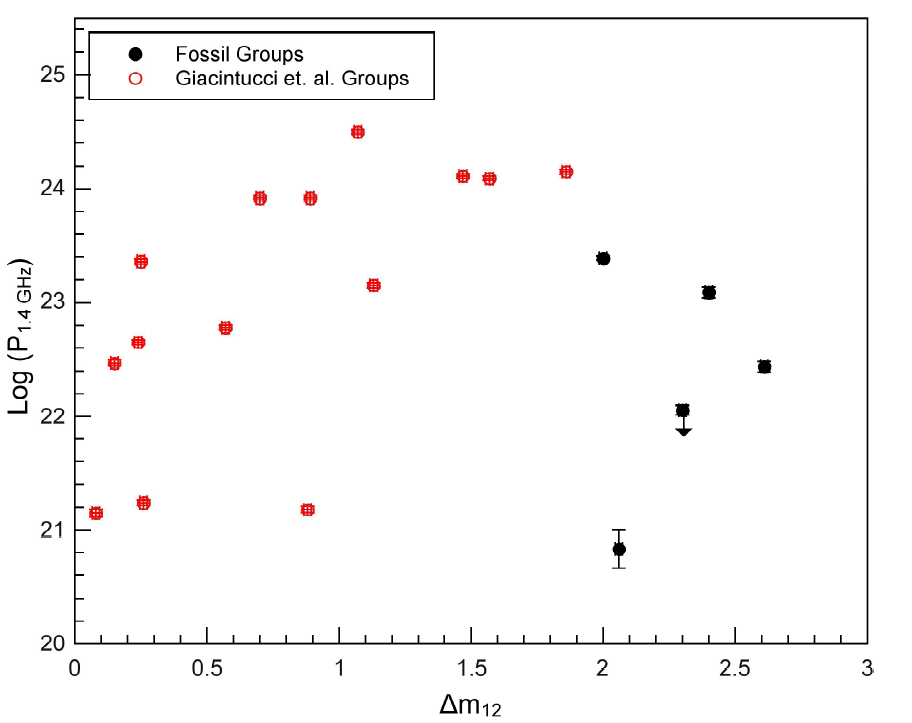,width=3.4in,height=3.2in}
\caption{1.4 GHz luminosity vs luminosity gap of
fossil galaxy groups (black) and Giacintucci et al. (2011)
groups (red).}
\label{figdm1420}
\end{figure}

\begin{figure}
\center
\epsfig{file=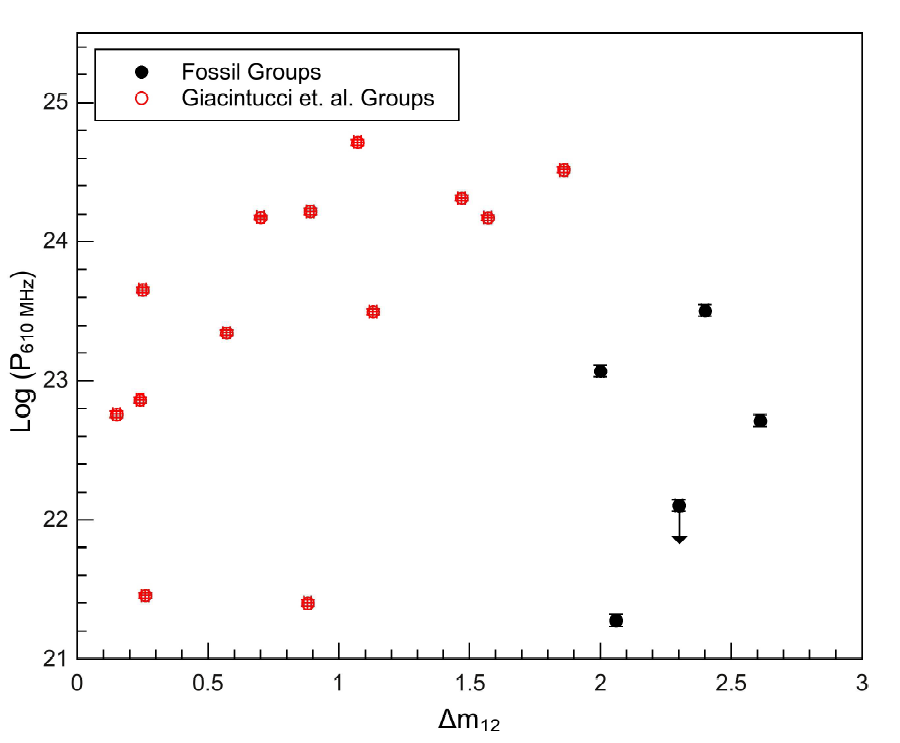,width=3.4in,height=3.2in}
\caption{610 MHz luminosity vs luminosity gap of
fossil galaxy groups (black) and Giacintucci et al. (2011)
groups (red).}
\label{figdm610}
\end{figure}

One of the goals of this study is to find out if the giant elliptical galaxies at the core of the fossil groups with large luminosity gap, are more, less or equally active in radio compared to the same in other galaxy groups. The radio luminosity vs. $K$-band luminosity of the BGGs at 1.4 GHz and 610 MHz is  shown in Figs. \ref{fig11} and \ref{fig12}, in which we also compare the fossil sample with a sample of  galaxy groups presented by \cite{Giacintucci11} . 
The control sample is the only sample of galaxy groups observed with GMRT at 610 MHz in the literature. The Ks luminosities in both the fossil and the comparison samples are extracted from the 2MASS extended source catalogue.  The 1.4 GHz luminosities in comparison sample are adopted from the NVSS catalogue. The best fit (Figs. \ref{fig11}-\ref{fig19}) is estimated based on the method of orthogonal regression \citep{Fuller87}, for each plot. J1331 appears to be the odd one in the sample. As mentioned before, it has a bright radio emission for a relatively dim BGG at ${K}_{s}$ band. Moreover, J1331 has inverse spectrum within the observing frequencies unlike other BGGs in this sample. We, thus, exclude this target from the fits. 

The BGGs in fossil groups appear to be under luminous for a given $K$-band luminosity,  as a proxy to the stellar mass, than the BGGs in the comparison groups. Thus it appears that despite a large stellar mass reservoir, the BGGs residing in fossil groups are currently less active. 
We can interpret this trend as a consequence of non-merging nature of fossil systems, as this appears to be the only major distinction between our sample and the comparison sample. In that case, the lack of recent galaxy merging lowers the efficiency of the AGN fuelling which results in less powerful AGN outburst.
 Alternatively, one could argue (Mulchaey and Zabludiff 1999) that fossils are drawn from an atypical luminosity function and thus have unusually high stellar masses. While we can not rule out this possibility, there are evidences which weaken this argument. Fossils have shown to contribute as large as 8-20 per cent to the population of galaxy clusters with the same X-ray luminosity. However, there are no studies in the literature which points out to such an unusual galaxy luminosity function from which fossils are formed. Furthermore, cosmological simulations show that fossil can form with the aforementioned abundance (Dariush et al 2007) and thus there is no need to treat them with any special mass function. 

The AGN radio power also scales with the total stellar mass budget of the group. This is presented as a correlation between the group total $K$-band luminosities and the AGN radio luminosity in Figs. \ref{fig13}  and \ref{fig14} . Group members have been determined from 
2MASS Redshift Survey (2MRS) group catalogue \citep{Crook07}. For some galaxy groups for which the data was not
available in the 2MRS group catalogue, we used \cite{Helsdon01} method to find group members in NED within half of the virial radius of galaxy groups, selecting the group members within a $\pm$0.002 redshift slice. The total $K$-band luminosity of comparison groups changes remarkably while for fossil groups with a bright BGG and few dim members, there is no significant change in the $K$-band luminosity. Figs. \ref{fig13}  and \ref{fig14} show that fossils fall comfortably within the distribution of the comparision sample.
Radio luminosities in fossil and non-fossil groups share the same scaling with the total stellar mass of the groups while their BGGs appear to present a significant offset (Figs. \ref{fig11} and \ref{fig12}), as discussed above. We will return to this in section \ref{sec:HIFLUGCS}.
The Spearman's rank correlation coefficient is estimated for all sample-plots in Figs. \ref{fig11}-\ref{fig14}, with 0.6-0.7 for 
comparison sample and 0.8-0.9 for fossil groups.

 In a similar attempt, \cite{Hess12} have found correlation between the radio 
power of BGGs and the luminosity gap ($\Delta m_{12}$) within the group.  Figs. \ref{figdm1420} and \ref{figdm610}  
show the radio power vs. the luminosity gap. We find no such correlation between the radio luminosity
and the luminosity gap in either of the observed frequencies. The radio luminosity of the comparison samples appears be correlated to 
its luminosity gap (Spearman's rank correlation coefficient of 0.6) but the trend is not followed by the fossil groups. 
In 610 MHz (Fig. \ref{figdm610}), the radio power of the fossil sample is statistically lower than the same in the comparison sample. Thus based on our observations we find no correlation between the radio power and the luminosity gap contrary to \cite{Hess12}. 
This means that the luminosity gap is not the only distinctive parameter in fossil-non fossil sample that decide for the correlation.
Other differences e.g. X-ray relaxed nature of fossils may play role in this trend.
\cite{Giacintucci11} selected elliptical dominated groups, possess structures in X-ray brightness and temperature distribution
or radio morphology, contrary to the fossil groups that they have relaxed and undisturbed X-ray emission.

\subsection {Radio luminosity and fossil galaxy groups characteristics  } 
\label{sec:HIFLUGCS}
Detection of the radio lobes in two fossil groups in our sample 
could either be the result of AGN duty cycle or it may have an origin in the group or host galaxy
properties. These two groups are the most X-ray luminous objects in the sample 
with largest cooling radius and highest outskirt temperature. Their BGGs are
brighter in $r$-band than the other fossil groups BGGs thus the probability that their BGGs 
host a radio loud AGN is also higher \citep{Best07}. Their bolometric radio luminosities are 
between 10 MHz to 15 GHz calculated using the spectral index listed in table 4, are also
the highest. \cite{Birzan04} shows that more powerful jets reside in more radio and X-ray 
luminous galaxies. Jet power is also tightly correlated to the total radio luminosity 
if one can eliminate the effect of break frequency \citep{Birzan08}.              
Accordingly, the radio bolometric luminosity is a measure for AGN power and
using this, we can investigate how large scale characteristics or central galaxy
properties control feedback mechanism in clusters and groups.

\begin{figure}
\epsfig{file=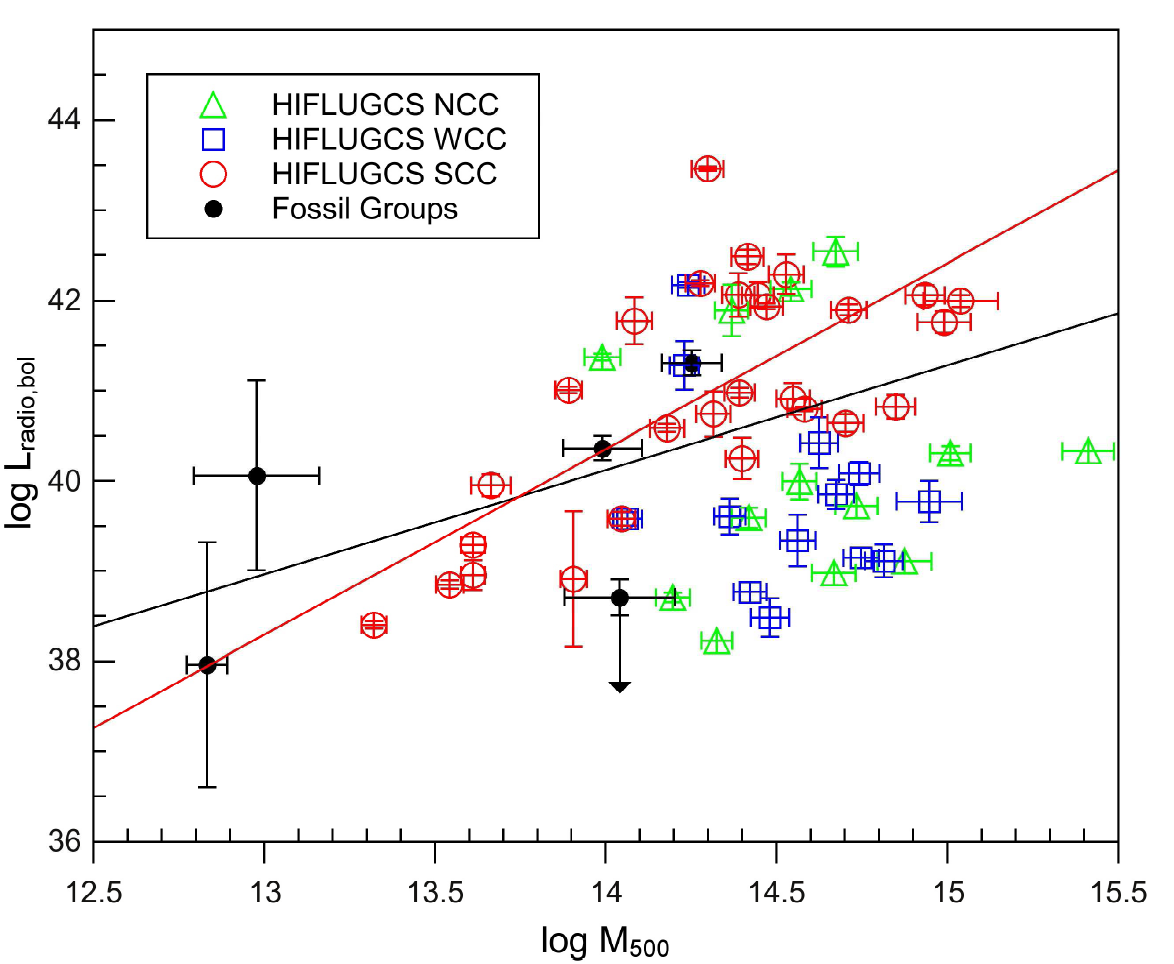,width=3.4in,height=3.0in}
\caption{Total radio luminosity vs virial mass for Strong Cool Core (SCC), Weak Cool Core (WCC) and Non Cool Core (NCC) clusters of HIFLUGCS 
 and fossil galaxy groups. The red and black lines correspond to fits to the red and black samples.}
\label{fig16}
\end{figure}

\begin{figure}
\epsfig{file=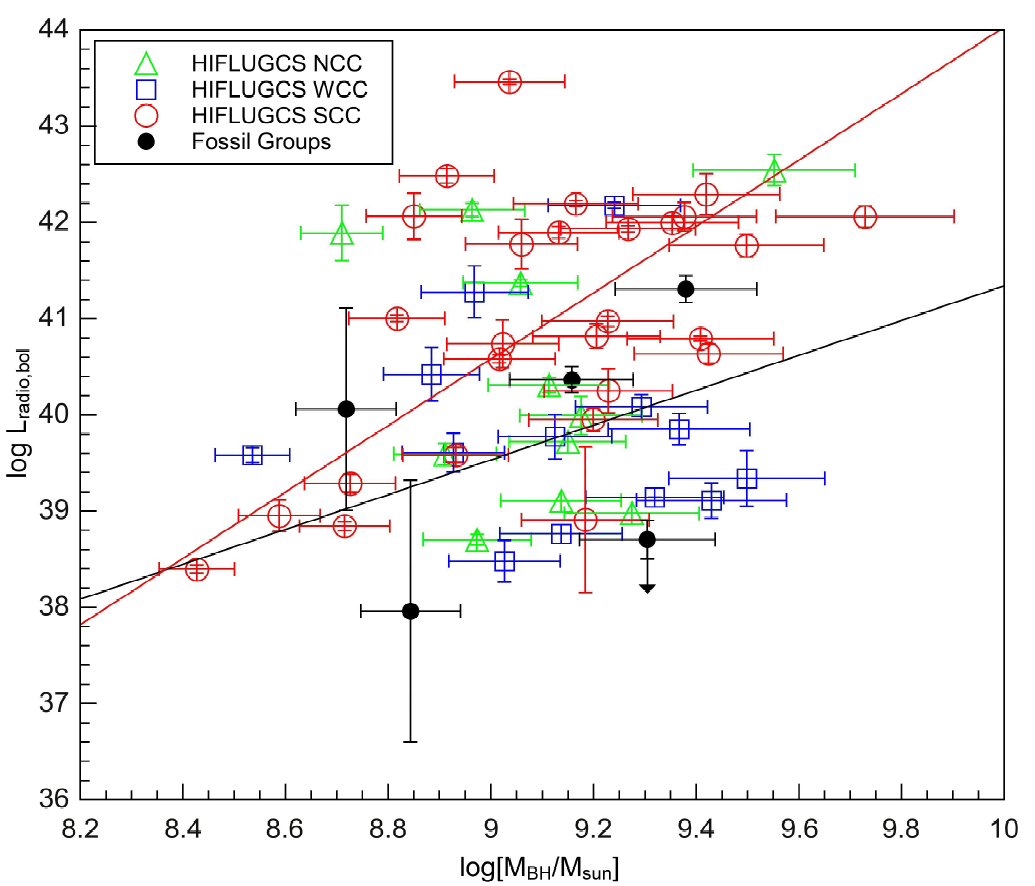,width=3.3in,height=2.9in}
\caption{Total radio luminosity vs black hole mass. The symbols are the same as Fig. \ref{fig16}.}
\label{fig17}
\end{figure}

\begin{figure}
\epsfig{file=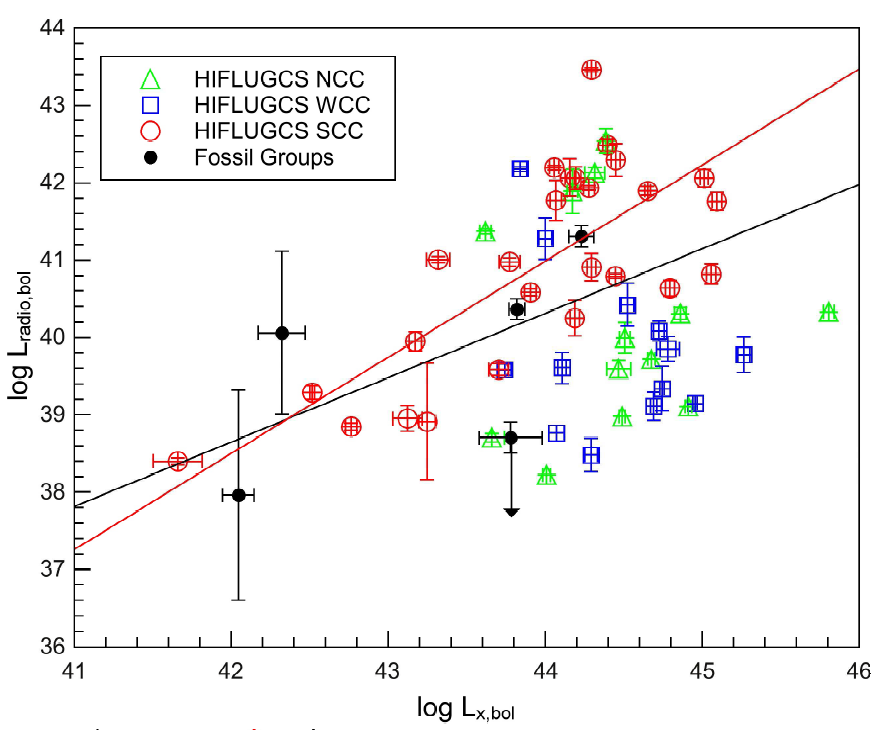,width=3.4in,height=3.0in}
\caption{Total radio luminosity vs X-ray luminosity. The symbols are the same as Fig. \ref{fig16}.}
\label{fig18}
\end{figure}

\begin{figure}
\epsfig{file=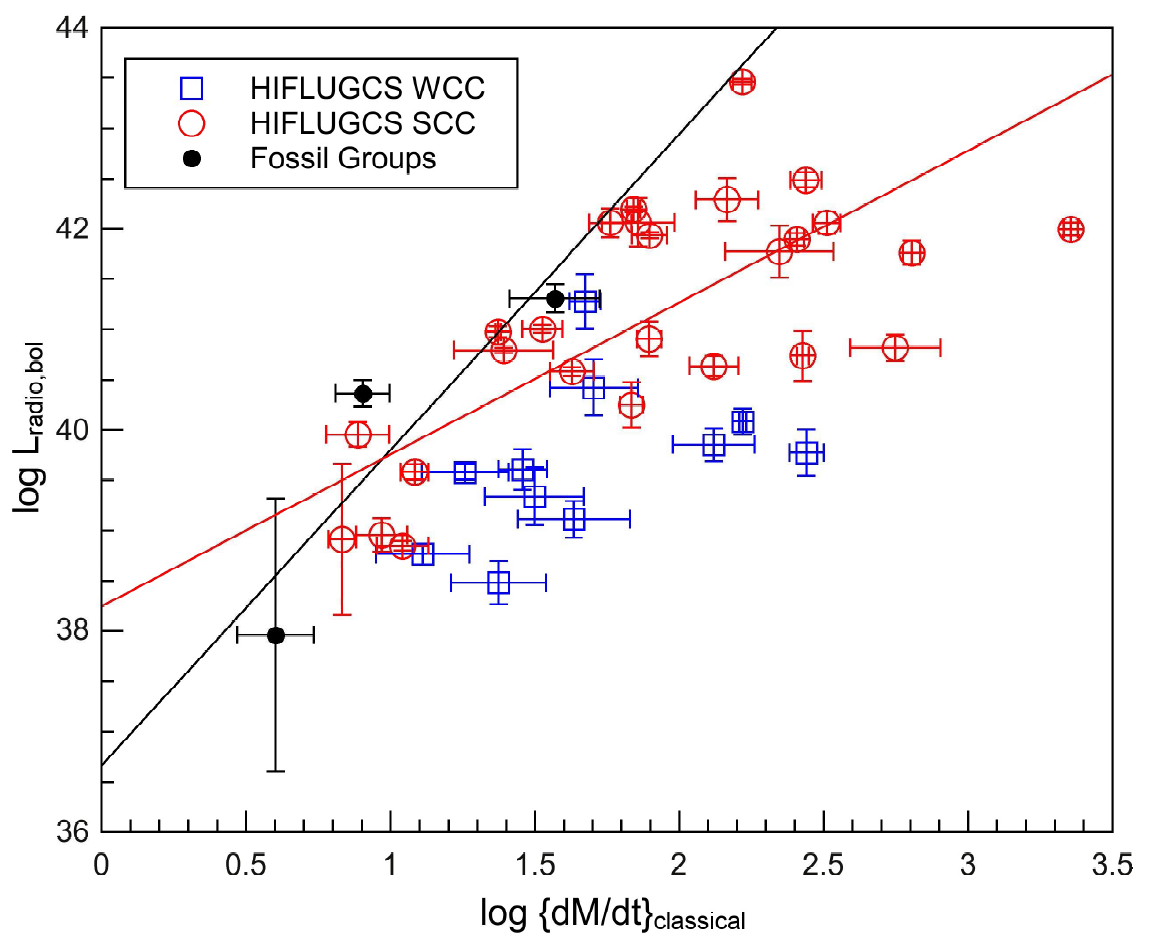,width=3.4in,height=3.0in}
\caption{Total radio luminosity vs classical mass deposition rate. The symbols are the same as Fig. \ref{fig16}.}
\label{fig19}
\end{figure}

\cite{Mittal09} shows, for HIFLUGCS sample of galaxy clusters, there is a correlation between virial mass, $M_{500}$, or classical mass deposition rate, $\dot{M}$, and the total radio luminosity in strong and weak cool core clusters as well as weak correlation between  black hole mass or X-ray luminosity and total radio luminosity in strong cool core clusters. In this section, we adapted HIFLUGCS as a control sample to compare 
with the fossil groups scaling relations. Figs. \ref{fig16}-\ref{fig19} show the results.
The total mass of the cluster, X-ray luminosity and the mass deposition rate are properties of the cluster and the IGM, while the black hole mass is a galaxy property.

Fossil galaxy groups are archetypal virialized and relaxed systems as they have a smooth and undisturbed 
X-ray emission map with their BGGs located at the position of the X-ray peak of the group. Such an unperturbed IGM is fertile to the formation a cool core. Strong cool cores are not detected in galaxy clusters with turbulent, unrelaxed, dynamically active and merging environments such as in Coma cluster (\cite{Simionescu13} and references therein). Sanderson et al (2009) have also shown that the offset between the BCGs and the X-ray centroid connected to the presence of cool cores, such that the groups with a large offset between the X-ray centroid and the BGG location lack cool cores. Moreover \cite{kpj07} showed that fossil groups have lower gas entropy than non-fossil groups specially for the low-mass fossils. This is consistent with the lower entropy observed at the SCC clusters against WCC and NCC \citep{Hudson10}.  Thus, one expects that fossils, given their relaxed nature, coincide with cool core clusters. Figs. \ref{fig16}, \ref{fig18} and \ref{fig19} show that at least  the distribution of the fossil groups is consistent with this argument. The large uncertainty in the radio luminosity of fossil groups sample prevents us from drawing a firm conclusion.

 Fossil galaxy groups harbour bright BCGs in optical and infrared wavelengths. Thus for a total optical luminosity of galaxy groups, and given the presence of a large luminosity gap, the share of the brightest group galaxies in the total optical luminosity is larger in fossil groups than in a non-fossil group, statistically. This is clearly seen in Figs. \ref{fig11}, \ref{fig12} and to some extend in Fig. \ref{fig17}. In contrast when the total optical luminosity of the group is taken into account, there is no distinction between fossil groups and non-fossil groups and the 1.4 GHz and 610 MHz radio luminosity of the BGG scales with the total stellar luminosity of the groups (Fig. \ref{fig13} and Fig. \ref{fig14}). As mentioned in section \ref{sec:OldAGN}, this behaviour might have an origin in special formation scenario for fossils or might simply indicate that the radio luminosity is more correlated with group and clusters properties than their BCGs.

We note that while generally the bolometric radio luminosities does not correlate to the clusters parameters, the concise classification of the sample into SCC, WCC and NCC according to the core cooling time shows that other parameter, in this case the cooling time, play a role. A lack of correlation between radio luminosity and cluster mass \citep{Stott12}, optical luminosity of BCGs \citep{Stott12} and AGNs \citep{Wadadekar04} are also reported.

\section{Summery and conclusions}
\label{sec:summery}

Fossil galaxy groups are archetypal virilised galaxy systems with smooth and undisturbed X-ray morphology, indicating the absence of a recent group scale merger, and a giant elliptical galaxy dominating optical luminosity of the group, that the brightest group galaxy has not experienced a recent galaxy major merger. They are thus ideal laboratories for studying IGM heating and cooling. We study a sample of 5 fossil galaxy groups at 610~MHz and 1.4~GHz using their GMRT observations. In two cases, ESO 3060170 and RX J1416.4+2315, radio lobes, as signs of recent AGN activities, are detected. We estimate heating rates via AGN mechanical, shock and vortex heating.

1- The mechanical power of AGNs are estimated by calculating the PdV on the IGM via radio lobes expansion and movement. We deduced a range of mechanical power corresponding to an overpressure factors of 2 and 10. The Heating power is not sufficient to keep IGM from cooling within $R_{c}$, in which the cooling time falls below the Hubble time, however, it is quit sufficient to account for the X-ray loss in the cool core radius, $R_{cc}$, where the temperature starts to fall towards the centre of the group. Radio bubbles in our sample migrate from the centre of the galaxy, up to the cool core radius, $R_{cc}$, thus powerful enough to stop their surrounding gas from cooling. Thus other sources of heating are necessary, specially in larger distances from the centre.    

2- The AGN shock power in ESO 3060170 is  $\sim$1$\times {10}^{42}$ erg {s}$^{-1}$ enough to heat the IGM within 
 $R_{cc}$ but still much less than the X-ray loss within the cooling radius.

3-Vortex heating is the least efficient as its heating power is an order of magnitude smaller than the amonut needed to quench cooling in the core even within the cool core radius, $R_{cc}$.

One could argue that the low efficiency of the AGN heating in our fossil sample galaxies could fundamentally be a consequence of lack of recent galaxy mergers with the giant elliptical galaxy in these systems.  In this argument, galaxy merger is viewed as an AGN fuelling mechanism which results in a more powerful cycles of AGN outburst. To probe this, we compared our sample with a sample of galaxy groups and clusters from \cite{Birzan04}. The results show mechanical powers in fossil and non-fossil objects do not differ significantly (Fig. \ref{fig15}). Nevertheless, drawing any firm conclusion requires a statistically larger sample of fossil galaxy groups.

Heat conduction and turbulent gas motions, induced by merger, AGN outburst or motions of galaxies \citep{Ruszkowski10} are among the candidates that can heat up IGM on their own and/or in combinations with AGNs mechanical heating. Unlike the AGN heating, these sources of energy can distribute the heat supply
symmetrically.  Our estimation of the heating through conduction at the centre of fossil galaxy group
ESO 3060170 shows that, neglecting magnetic field, this source of energy is ten times more powerful to heat up cool core but poorly understood magnetic field still leaves the challenge unsolved.

We studied feeding mechanisms of super massive black hole. For ESO 3060170, the Bondi power is larger than AGN power. Spin powered feedback is also possible 
since a rotating black hole with spin parameter a = 0.002 can send out jets as energetic 
as the one observed in ESO 3060170 and RX J1416.4+2315.  

The 1.4 GHz and the 610 MHz luminosity of fossil sample were compared with a non-fossil group sample
\citep{Giacintucci11}, for a given total stellar mass of the group and also for a given stellar mass of the brightest group galaxy. The results (Figs. \ref{fig11}- \ref{fig14}) demonstrate that the fossil brightest group galaxies are under luminous in radio luminosity, for a given $K$-band luminosity of the dominant galaxy. However, when the total stellar mass of the group is taken into account, fossil and non-fossil samples are inseparable in the $L_{radio}$-$L_{k,group}$ plane. About 40$ \pm $28 per cent of fossils and 44$ \pm$ 16 per cent of \cite{Giacintucci11} groups are radio loud ($L_{radio} > 10^{23}$ W {Hz}$^{-1}$). 
We also compared total radio luminosity of fossil galaxy groups with the same from HIFLUGCS 
for a given black hole mass, classical mass deposition rate, X-ray bolometric luminosity 
and the group halo mass ($M_{500}$).  While there is a large statistical uncertainties given the size of the sample, it appears that the distribution of fossil galaxy groups, which we except to be ideal environments for the formation of the cool cores, is not inconsistent with the distribution of the strong cool core clusters in various scaling relations based on the above properties of clusters.

Our results suggest that group property has direct impact on the radio outburst of the 
AGN and thus the radio luminosity of a giant elliptical galaxy is environment dependent. 
We already know that a higher fraction of radio loud AGNs are observed in galaxy clusters and groups rather than in the field galaxies \citep{Best07,Lin07,Stott12}, however, the field galaxies are normally not very massive or optically luminous. In this contribution we show that fossil dominant galaxies, despite their high luminosity, are relatively less radio active.  In Figs. \ref{fig11} and \ref{fig12} we show that fossil dominant galaxies are strikingly under luminous in radio. Figs. \ref{fig16} through \ref{fig19} take this to a next level arguably showing that the core of the fossil groups is subject to a strong cooling at the same time their central AGN is less radio active than one could have expected from their stellar masses. 

 There are evidences that the brightest group galaxies have non-boxy isophotes \citep{kpj06, Smith10}, which is consistent with the picture in which these giant elliptical galaxies are formed via multiple gas rich mergers early in their formation history. In other words, they are not a dry merger product. This has been challenged by a more recent study \citep{Barbera12} which find mostly boxy isophotes for a sample of fossil BGGs. The study of \citet{Smith10} is based on Hubble Space Telescope observations and is the largest sample to date which suggests a connection between the luminosity gap and the nature of the isophotal shape of the BGGs. The sample studied here largely overlaps with the sample of \cite{kpj06} which finds the majority of the sample BGGs to have non-boxy isophotes and as shown in this study they are relatively dimmer in the radio emission. While the shape of the isophotes can be used to argue for the lack of recent dry majors, minor mergers can continue to occur since the epoch of the major merger. If so, it appears that they do not contribute significantly to the feeding of the central black hole, as compared to the BGGs in general population of the groups. 

\section*{Acknowledgments}
The authors thank Max Planck Institute for Radio Astronomy, Bonn, Germany and 
National Centre for Radio Astrophysics, Pune, India  for their hospitality.
H.Miraghaei would like to thanks Dr James Anderson and Dr Thomas Reiprich for their suggestions
during the data analysis, the Iranian National Observatory Project and the National Elite Foundation, Iran, for partial financial support.

\label{lastpage}

\begin{thebibliography}{99}

\bibitem[\protect\citeauthoryear{Allen \etal}{2006}]{Allen06}
Allen S. W., Dunn R. J. H., Fabian A. C., Taylor G. B., Reynolds C. S., 2006 , MNRAS, 372, 21	
\bibitem[\protect\citeauthoryear{Birzan \etal}{2008}]{Birzan08}
B\^\i rzan L., McNamara B. R., Nulsen P. E. J., Carilli C. L., Wise M. W., 2008, ApJ,
 686, 859
\bibitem[\protect\citeauthoryear{Birzan \etal}{2004}]{Birzan04}
B\^\i rzan L., Rafferty D. A., McNamara B. R., Wise M. W., Nulsen P. E. J., 2004, ApJ, 607, 800
\bibitem[\protect\citeauthoryear{Baldi \etal}{2009}]{Baldi09}	
Baldi A., Forman W., Jones C., Nulsen P., David L., Kraft R., Simionescu A., 2009, ApJ, 694, 479
\bibitem[\protect\citeauthoryear{Begelman}{2001}]{Begelman01}
Begelman M. C., in Astronomical Society of the Pacific Conference Series, Vol. 240, Gas and Galaxy Evolution, ed.
J. E. Hibbard, M. Rupen, J. H. van Gorkom, 363
\bibitem[\protect\citeauthoryear{Best \etal}{2007}]{Best07}	
Best P. N., von der Linden A., Kauffmann G., Heckman T. M., Kaiser C. R., 2007,
MNRAS, 379, 894
\bibitem[\protect\citeauthoryear{Blanton \etal}{2010}]{Blanton10}
Blanton E. L., Clarke T. E., Sarazin C. L., Randall S. W., McNamara B. R., 
2010, Publ. of the Natl. Acad. of Sci., 107, 7174
\bibitem[\protect\citeauthoryear{Blanton \etal}{2009}]{Blanton09}
Blanton E. L., Randall S. W., Douglass E. M., Sarazin C. L., Clarke T. E., McNamara B. R., 2009, ApJ, 697, 95
\bibitem[\protect\citeauthoryear{Bondi}{1952}]{Bondi52}	
Bondi H., 1952, MNRAS, 112, 195
\bibitem[\protect\citeauthoryear{Bornancini \etal}{2010}]{Bornancini10}
Bornancini C. G., O'Mill A. L., Gurovich S., Lambas D. G., 2010, MNRAS, 406, 197
\bibitem[\protect\citeauthoryear{Bregman \& David}{1988}]{Bregman88}
Bregman J. N., David L. P., 1988, ApJ, 326, 639 
\bibitem[\protect\citeauthoryear{Brown}{1987}]{Brown87}
Brown R. L., 1987, in Spectroscopy of astrophysical plasmas, 35
\bibitem[\protect\citeauthoryear{Cavagnolo \etal}{2010}]{Cavagnolo10}
Cavagnolo K. W., McNamara B. R., Nulsen P. E. J., Carilli C. L., Jones C., B\^\i rzan L., ApJ, 
720, 1066
\bibitem[\protect\citeauthoryear{Churazov \etal}{2002}]{Churazov02}
Churazov E., Sunyaev R., Forman W., B\"{o}hringer H., 2002, MNRAS, 332, 729
\bibitem[\protect\citeauthoryear{Clarke \etal}{2009}]{Clarke09}
Clarke T. E., Blanton E. L., Sarazin C. L., Anderson L. D., Gopal-Krishna, Douglass E. M., Kassim N. E., 2009, ApJ, 697, 1481
\bibitem[\protect\citeauthoryear{Crook \etal}{2007}]{Crook07}
Crook A. C., Huchra J. P., Martimbeau N.ie., Masters K. L., Jarrett T., Macri L. M., 2007,
ApJ, 655, 790
\bibitem[\protect\citeauthoryear{Dariush \etal}{2007}]{dariush07}
Dariush A. A., Khosroshahi H. G., Ponman T. J., Pearce F., Raychaudhury S., Hartly W., 2007, MNRAS, 382, 433 
\bibitem[\protect\citeauthoryear{Dariush \etal}{2010}]{dariush10}	
Dariush A. A., Raychaudhury S., Ponman T. J., Khosroshahi H. G., Benson A. J., Bower R. G., Pearce F., 2010, MNRAS, 405, 1873
\bibitem[\protect\citeauthoryear{Domainko \etal}{2004}]{Domainko04}
Domainko W., Gitti M., Schindler S., Kapferer W., 2004, A\&A, 425, 21
\bibitem[\protect\citeauthoryear{Dunn \etal}{2010}]{Dunn10}
Dunn R. J. H., Allen S. W., Taylor G. B., Shurkin K. F., Gentile G., Fabian A. C., Reynolds C. S.,MNRAS, 404, 180
\bibitem[\protect\citeauthoryear{Dunn \& Fabian}{2006}]{DunnF06}	
Dunn R. J. H., Fabian A. C., 2006, MNRAS, 373, 959
\bibitem[\protect\citeauthoryear{Dunn \& Fabian}{2004}]{Dunn04}
Dunn R. J. H., Fabian A. C., 2004, MNRAS, 355, 862
\bibitem[\protect\citeauthoryear{Dunn, Fabian \& Taylor}{2005}]{Dunn05}
Dunn R. J. H., Fabian A. C., Taylor G. B., 2005, MNRAS, 364, 1343
\bibitem[\protect\citeauthoryear{Eigenthaler \& Zeilinger}{2009}]{Eigenthaler09}	
Eigenthaler P., Zeilinger W. W., 2009, AN, 330, 978
\bibitem[\protect\citeauthoryear{Eilek \& Weatherall}{1999}]{Eilek99}
Eilek J., Weatherall J. C., 1999, in Diffuse thermal and relativistic plasma 
in galaxy clusters, ed. H. B\"{o}hringer, L. Feretti, P. Schuecker., ( Garching: MPE Rept. 271 ), 249
\bibitem[\protect\citeauthoryear{Ellison \etal}{2011}]{Ellison11}
Ellison S. L., Patton D. R., Mendel J. T., Scudder J. M., 2011,	MNRAS, 418, 2043
\bibitem[\protect\citeauthoryear{Fabian \etal}{2003}]{Fabian03}	
Fabian A. C., Sanders J. S., Allen S. W., Crawford C. S., Iwasawa K., Johnstone R. M., Schmidt R. W., Taylor G. B.,
2003, MNRAS Letters, 344, 43
\bibitem[\protect\citeauthoryear{Fabian \etal}{2006}]{Fabian06}
Fabian A. C., Sanders J. S., Taylor G. B., Allen S. W., Crawford C. S., Johnstone R. M., Iwasawa K., 2006
MNRAS, 366, 417
\bibitem[\protect\citeauthoryear{Feretti \& Giovannini}{2008}]{Feretti08}
Feretti L., Giovannini G., 2008 in A Pan-Chromatic View of Clusters of Galaxies and 
the Large-Scale Structure, ed. M. Plionis, O. L\'opez-Cruz, D. Hughes, Lect. Notes Phys., 740, 143 (Springer)
\bibitem[\protect\citeauthoryear{Feretti \etal}{2012}]{Feretti12}
Feretti L., Giovannini G., Govoni F., Murgia M., 2012, A\&ARv, 20, 54
\bibitem[\protect\citeauthoryear{Forman \etal}{2007}]{Forman07}
Forman W., Jones C., Churazov E., Markevitch M., Nulsen P., Vikhlinin A., Begelman M., 
B\"{o}hringer H., Eilek J., Heinz S., Kraft R., Owen, F., Pahre M., 2007, ApJ, 665, 1057
\bibitem[\protect\citeauthoryear{Friedman, Heinz \& Churazov}{2012}]{Friedman12}	
Friedman S. H., Heinz S., Churazov E., 2012, ApJ, 746, 112
\bibitem[\protect\citeauthoryear{Fuller}{1987}]{Fuller87}
Fuller W. A., Measurement error models, Wiley Series in Probability and Mathematical Statistics, New York: Wiley, 1987
\bibitem[\protect\citeauthoryear{Gomez \etal}{2002}]{Gomez02}	
G\'omez P. L., Loken C., Roettiger K., Burns J. O, 2002, ApJ, 569, 122
\bibitem[\protect\citeauthoryear{Giacintucci \etal}{2011}]{Giacintucci11}		
Giacintucci S., O'Sullivan E., Vrtilek J., David L. P., Raychaudhury S., Venturi T., 
Athreya R. M., Clarke T. E., Murgia M., Mazzotta P., Gitti M., Ponman T., 
Ishwara-Chandra C. H., Jones C., Forman W. R., 2011, ApJ, 732, 95 
\bibitem[\protect\citeauthoryear{Gitti \etal}{2012}]{Gitti12}	
Gitti M., Brighenti F., McNamara B. R., 2012,
Advances in Astronomy, 2012, 1
\bibitem[\protect\citeauthoryear{Gitti \etal}{2010}]{Gitti10}
Gitti M., O'Sullivan E., Giacintucci S., David L. P., Vrtilek J., Raychaudhury S., Nulsen P. E. J., 
2010, ApJ, 714, 758
\bibitem[\protect\citeauthoryear{Guo \& Oh}{2008}]{Guoh08}
Guo F., Oh S. P., 2008, MNRAS, 384, 251
\bibitem[\protect\citeauthoryear{Guo, Oh \& Ruszkowski}{2008}]{Guo08}
Guo F., Oh S. P., Ruszkowski M., 2008, ApJ, 688, 859
\bibitem[\protect\citeauthoryear{Heinz \etal}{2006}]{Heinz06}
Heinz S., Br\"uggen M., Young A., Levesque E., 2006, MNRAS Letters, 373, 65
\bibitem[\protect\citeauthoryear{Helsdon \etal}{2001}]{Helsdon01}
Helsdon S. F., Ponman T. J., O'Sullivan E., Forbes D. A.. 2001,	
MNRAS, 325, 693
\bibitem[\protect\citeauthoryear{Helsdon \& Ponman}{2000}]{Helsdon00}
Helsdon S. F., Ponman T. J., 2000, MNRAS, 315, 356
\bibitem[\protect\citeauthoryear{Hess, Wilcots \& Hartwick}{2012}]{Hess12}
Hess K. M., Wilcots E. M., Hartwick V. L., 2012, AJ, 144, 48
\bibitem[\protect\citeauthoryear{Hudson \etal}{2010}]{Hudson10}
Hudson D. S., Mittal R., Reiprich T. H., Nulsen P. E. J., Andernach H., Sarazin C. L., 2010, 
A\&A, 513, 37
\bibitem[\protect\citeauthoryear{Jetha \etal}{2008}]{Jetha08}	
Jetha N. N., Hardcastle M. J., Babul A., O'Sullivan E., Ponman T. J., Raychaudhury S., Vrtilek J.,
 2008, MNRAS, 384, 1344
\bibitem[\protect\citeauthoryear{Jetha \etal}{2009}]{Jetha09}
Jetha N. N., Khosroshahi H., Raychaudhury S., Sengupta C., Hardcastle M., 2009,
AIP Conf. Proc., 1201, 305
\bibitem[\protect\citeauthoryear{Jones \etal}{2003}]{Jones03}	
Jones L. R., Ponman T. J., Horton A., Babul A., Ebeling H., Burke D. J., 2003, MNRAS, 343, 627
\bibitem[\protect\citeauthoryear{Jones \etal}{1998}]{Jones98}	
Jones L. R., Scharf C., Ebeling H., Perlman E., Wegner G., Malkan M., Horner D., 1998, ApJ, 495, 100
\bibitem[\protect\citeauthoryear{Jones \etal}{2000}]{Jones00}	
Jones D. L., Wehrle A. E., Meier D. L., Piner B. G., 2000, ApJ, 534, 165
\bibitem[\protect\citeauthoryear{Kellermann \& Pauliny-Toth}{1981}]{Kellermann81}
Kellermann K. I., Pauliny-Toth I. I. K., 1981, ARA\&A, 19, 373
\bibitem[\protect\citeauthoryear{Kellermann, Pauliny-Toth \& Williams}{1969}]{Kellermann69}	
Kellermann K. I., Pauliny-Toth I. I. K., Williams P. J. S., 1969, ApJ, 157, 1 
\bibitem[\protect\citeauthoryear{Kempner, Sarazin \& Markevitch}{2006}]{Kempner03}
Kempner J. C., Sarazin C. L., Markevitch, M., 2003, ApJ, 593, 291
\bibitem[\protect\citeauthoryear{Khosroshahi, Jones \& Ponman}{2004}]{kjp04}
Khosroshahi H. G., Jones L. R. and Ponman T. J., 2004, MNRAS, 349, 1240
\bibitem[\protect\citeauthoryear{Khosroshahi \etal}{2006}]{kmpj06}
Khosroshahi H. G., Maughan B., Ponman T. J., and Jones L. R., 2006, 
MNRAS, 369, 1211	
\bibitem[\protect\citeauthoryear{Khosroshahi, Ponman \& Jones}{2006}]{kpj06}
Khosroshahi H. G., Ponman T. J., and Jones L. R., 2006, MNRAS Letters, 372, 68
\bibitem[\protect\citeauthoryear{Khosroshahi, Ponman \& Jones}{2007}]{kpj07}
Khosroshahi H. G., Ponman T. J., and Jones L. R., 2007, MNRAS, 377, 595
\bibitem[\protect\citeauthoryear{Kim \& Narayan}{2003}]{Kim03}
Kim W.-T., Narayan R., 2003, ApJ, 596, 889
\bibitem[\protect\citeauthoryear{La Barbera \etal}{2009}]{Barbera09}		
La Barbera F., de Carvalho R. R., de la Rosa I. G., Sorrentino G., Gal R. R., Kohl-Moreira J. L., 2009, 
AJ, 137, 3942 
\bibitem[\protect\citeauthoryear{La Barbera \etal}{2012}]{Barbera12}		
La Barbera F., Paolillo, M., De Filippis, E., de Carvalho, R. R, 2012, MNRAS, 422, 3010 
\bibitem[\protect\citeauthoryear{Laing \& Peacock}{1980}]{Laing80} 
Laing R. A., Peacock J. A., 1980, MNRAS, 190, 903
\bibitem[\protect\citeauthoryear{Levinson, Laor \& Vermeulen}{1995}]{Levinson95}
Levinson A., Laor A., Vermeulen R. C., 1995, ApJ, 448, 589
\bibitem[\protect\citeauthoryear{Lin \& Mohr}{2012}]{Lin07}	
Lin Y., Mohr J. J., 2007, ApJS, 170, 71
\bibitem[\protect\citeauthoryear{Lopes de Oliveira \etal}{2010}]{Lopes10}	
Lopes de Oliveira R., Carrasco E. R., Mendes de Oliveira C., Bortoletto D. R., Cypriano E., Sodr\'e L. Jr., Lima Neto G. B.,
 2010, AJ, 139, 216
\bibitem[\protect\citeauthoryear{Marconi \& Hunt}{2003}]{Marconi03}
Marconi A., Hunt L. K., 2003, ApJ, 589, 21
\bibitem[\protect\citeauthoryear{Markevitch, Sarazin \& Vikhlinin}{1999}]{Markevitch99}
Markevitch M., Sarazin C. L., Vikhlinin A., 1999, ApJ, 521, 526
\bibitem[\protect\citeauthoryear{Markevitch \& Vikhlinin}{2007}]{Markevitch07}
Markevitch M., Vikhlinin A., 2007, PhR, 443, 1
\bibitem[\protect\citeauthoryear{Mauch \etal}{2013}]{Mauch13}
Mauch T., Klöckner H.-R., Rawlings S., Jarvis M. J., Hardcastle M. J., Obreschkow D., Saikia D. J., Thompson M. A.,
2013, Accepted by MNRAS, eprint arXiv:1307.4590
\bibitem[\protect\citeauthoryear{McCarthy \etal}{2010}]{McCarthy10}	
McCarthy I. G., Schaye J., Ponman T. J., Bower R. G., Booth C. M., Dalla Vecchia C., 
Crain R. A., Springel V., Theuns T., Wiersma R. P. C., 2010, MNRAS, 406, 822
\bibitem[\protect\citeauthoryear{McNamara \& Nulsen}{2007}]{McNamara07}
McNamara B. R., Nulsen P. E. J., 2007, ARA\&A, 45, 117
\bibitem[\protect\citeauthoryear{McNamara, Rohanizadegan \& Nulsen}{2011}]{McNamara11}
McNamara B. R., Rohanizadegan M., Nulsen P. E. J., 2011, ApJ, 727, 39 
\bibitem[\protect\citeauthoryear{McNamara \etal}{2005}]{McNamara05}	
McNamara B. R., Nulsen P. E. J., Wise M. W., Rafferty D. A., Carilli C., Sarazin C. L., Blanton E. L.,
2005, Nature, 433, 45
\bibitem[\protect\citeauthoryear{Megn}{2008}]{Megn08}
Megn A. V, 2008, Astronomy Reports, 52, 201
\bibitem[\protect\citeauthoryear{Meier}{1999}]{Meier99}		
Meier D. L., 1999, ApJ, 522, 753
\bibitem[\protect\citeauthoryear{Mendes de Oliveira, Cypriano \& Sodré}{2006}]{Mendes06}
Mendes de Oliveira C. L., Cypriano E. S., Sodr\'e L. Jr., 2006, AJ, 131, 158
\bibitem[\protect\citeauthoryear{Miller \etal}{2012}]{Miller12}	
Miller E. D., Rykoff E. S., Dupke R. A., Mendes de Oliveira C., Lopes de Oliveira R., 
Proctor R. N., Garmire G. P., Koester B. P., McKay T. A.,
 2012, ApJ, 747, 94
\bibitem[\protect\citeauthoryear{Million \etal}{2010}]{Million10}
Million E. T., Werner N., Simionescu A., Allen S. W., Nulsen P. E. J., Fabian A. C., B\"{o}hringer H., Sanders J. S.,
 2010, MNRAS, 407, 2046
\bibitem[\protect\citeauthoryear{Miraghaei \etal}{2014}]{miraghaei12}
Miraghaei H., Khosroshahi H. G.,  Sengupta C., Raychaudhury S., Jetha N. N., Abbassi S., Submitted. 
\bibitem[\protect\citeauthoryear{Mittal \etal}{2009}]{Mittal09}	
Mittal R., Hudson D. S., Reiprich T. H., Clarke T., 2009, A\&A, 501, 835
\bibitem[\protect\citeauthoryear{Motl \etal}{2004}]{Motl04}	
Motl P. M., Burns J. O., Loken C., Norman M. L., Bryan G., 2004, ApJ, 606, 635
\bibitem[\protect\citeauthoryear{Narayan \& Medvedev}{2001}]{Narayan01}
Narayan R., Medvedev M. V., 2001, ApJ, 562, 129
\bibitem[\protect\citeauthoryear{Nemmen \etal}{2007}]{Nemmen07}
Nemmen R. S., Bower R. G., Babul A., Storchi-Bergmann T., 2007, MNRAS, 377, 1652
\bibitem[\protect\citeauthoryear{Nulsen \etal}{2007}]{Nulsen07}
Nulsen P. E. J., Jones C., Forman W. R., David L. P., McNamara B. R., Rafferty D. A., B\^\i rzan L., Wise M. W., 2007,
in Heating versus Cooling in Galaxies and Clusters of Galaxies, 210
\bibitem[\protect\citeauthoryear{Nulsen \etal}{2005}]{Nulsen05}	
Nulsen P. E. J., McNamara B. R., Wise M. W., David L. P., 2005
ApJ, 628, 629
\bibitem[\protect\citeauthoryear{Nusser, Silk \& Babul}{2006}]{Nusser06}	
Nusser A., Silk J., Babul A., 2006, MNRAS, 373, 739
\bibitem[\protect\citeauthoryear{O'Sullivan \etal}{2012}]{OSullivan12}	
O'Sullivan E., Giacintucci S., Babul A., Raychaudhury S., Venturi T., Bildfell C.,
 Mahdavi A., Oonk J. B. R., Murray N., Hoekstra H., Donahue M., 2012 
MNRAS, 424, 2971
\bibitem[\protect\citeauthoryear{O'Sullivan \etal}{2011}]{OSullivan11}	
O'Sullivan E., Giacintucci S., David L. P., Gitti M., Vrtilek J. M., Raychaudhury S., Ponman T. J., 2011
ApJ, 735, 11
\bibitem[\protect\citeauthoryear{Okamoto, Nemmen \& Bower}{2008}]{Okamoto08}
Okamoto T., Nemmen R. S., Bower R. G., 2008, MNRAS, 385, 161
\bibitem[\protect\citeauthoryear{Osmond \& Ponman}{2004}]{Osmond04}
Osmond J. P. F., Ponman T. J., 2004, MNRAS, 350, 1511
\bibitem[\protect\citeauthoryear{Pacholczyk}{1970}]{Pacholczyk70}
Pacholczyk A. G., 1970, Radio astrophysics. Nonthermal processes in galactic and extragalactic sources
\bibitem[\protect\citeauthoryear{Perlman \etal}{2002}]{Perlman02}	
Perlman E. S., Horner D. J., Jones L. R., Scharf C. A., Ebeling H., Wegner G., Malkan M., 2002, ApJS, 140, 265
\bibitem[\protect\citeauthoryear{Peterson \& Fabian}{2006}]{Peterson06}
Peterson J. R., Fabian A. C., 2006, Physics Reports, 427, 1
\bibitem[\protect\citeauthoryear{Pierini \etal}{2011}]{Pierini11}
Pierini D., Giodini S., Finoguenov A., B\"ohringer H., D'Onghia E., Pratt G. W., D\'emocl\'es J., 
Pannella M., Zibetti S., Braglia F. G., Verdugo M., Ziparo F., Koekemoer A. M., 
Salvato M. \& COSMOS Collaboration, MNRAS, 417, 2927
\bibitem[\protect\citeauthoryear{Pizzolato \& Soker}{2012}]{Pizzolato10}	
Pizzolato F., Soker N., 2010, MNRAS, 408, 961
\bibitem[\protect\citeauthoryear{Ponman \etal}{1994}]{Ponman94}
Ponman T. J., Allan D. J., Jones L. R., Merrifield M., McHardy I. M., Lehto H. J., Luppino G. A., 1994, 
Nature, 369, 462
\bibitem[\protect\citeauthoryear{Pope, Mendel \& Shabala}{2012}]{Pope12}		
Pope E. C. D., Mendel J. T., Shabala S. S., 2012, MNRAS, 419, 50
\bibitem[\protect\citeauthoryear{Proctor \etal}{2011}]{Proctor11}	
Proctor R. N., de Oliveira C. M., Dupke R., de Oliveira R. L., Cypriano E. S., Miller E. D., Rykoff E.,
 2011, MNRAS, 418, 2054
\bibitem[\protect\citeauthoryear{Rafferty \etal}{2006}]{Rafferty06}
Rafferty D. A., McNamara B. R., Nulsen P. E. J., Wise M. W., 2006, ApJ, 652, 216
\bibitem[\protect\citeauthoryear{Randall \etal}{2011}]{Randall11}	
Randall S. W., Forman W. R., Giacintucci S., Nulsen P. E. J., Sun M., Jones C., Churazov E., David L. P., Kraft R., Donahue M., Blanton E. L., Simionescu A., Werner N., 2011, ApJ, 726, 18
\bibitem[\protect\citeauthoryear{Randall \etal}{2009}]{Randall09}
Randall S. W., Jones C., Markevitch M., Blanton E. L., Nulsen P. E. J., Forman W. R., 2009, ApJ, 700, 1404
\bibitem[\protect\citeauthoryear{Reiprich \& b\"{o}hringer}{2002}]{Reiprich02}
Reiprich T. H., B\"{o}hringer H., 2002, ApJ, 567, 716
\bibitem[\protect\citeauthoryear{Ruszkowski \& Begelman}{2002}]{Ruszkowski02} 
Ruszkowski M., Begelman M. C., 2002, ApJ, 581, 223
\bibitem[\protect\citeauthoryear{Ruszkowski \& Oh}{2011}]{Ruszkowski11} 
Ruszkowski M., Oh S. P., 2011, MNRAS, 414, 1493
\bibitem[\protect\citeauthoryear{Ruszkowski \& Oh}{2010}]{Ruszkowski10}
Ruszkowski M., Oh S. P., 2010, ApJ, 713, 1332
\bibitem[\protect\citeauthoryear{Sanders, Fabian \& Taylor}{2009}]{Sanders09}
Sanders J. S., Fabian A. C., Taylor G. B., 2009, MNRAS, 393, 71
\bibitem[\protect\citeauthoryear{Sanderson, Edge \& Smith}{2009}]{Sanderson09}
Sanderson A. J. R., Edge A. C., Smith G. P., 2009, MNRAS, 398, 1698
\bibitem[\protect\citeauthoryear{Santos, Mendes de Oliveira \& Sodré}{2007}]{Santos07}	
Santos W. A., Mendes de Oliveira C., Sodr\'e L. Jr., 2007,
 AJ, 134, 1551
\bibitem[\protect\citeauthoryear{Sarazin}{2002}]{Sarazin02}
Sarazin C. L., 2002, in Merging Processes in Galaxy Clusters, eds. L. Feretti, I.M. Gioia, G. Giovannini,
272, 1
\bibitem[\protect\citeauthoryear{Sarazin}{1988}]{Sarazin88}
Sarazin C. L., 1988, X-Ray Emission from Clusters of Galaxies, Cambridge Astrophysics Series, Cambridge: Cambridge University Press
\bibitem[\protect\citeauthoryear{Scharf \etal}{1997}]{Scharf97}	
Scharf C. A., Jones L. R., Ebeling H., Perlman E., Malkan M., Wegner G., 1997, ApJ, 477, 79
\bibitem[\protect\citeauthoryear{Schirmer \etal}{2010}]{Schirmer10}	
Schirmer M., Suyu S., Schrabback T., Hildebrandt H., Erben T., Halkola A., 2010, A\&A, 514, 60
\bibitem[\protect\citeauthoryear{Scott \& Readhead}{1977}]{Scott77}
Scott M. A., Readhead A. C. S., 1977, MNRAS, 180, 539
\bibitem[\protect\citeauthoryear{Sijacki \& Springel}{2006}]{Sijacki06}
Sijacki D., Springel V., 2006, MNRAS, 366, 397
\bibitem[\protect\citeauthoryear{Simionescu \etal}{2013}]{Simionescu13}	
Simionescu A., Werner N., Urban O., Allen S. W., Fabian A. C., Mantz A., Matsushita K., Nulsen P. E. J., Sanders J. S., Sasaki T., Sato T.,
 Takei Y., Walker S. A., submitted to ApJ, eprint arXiv:1302.4140
\bibitem[\protect\citeauthoryear{Smith \etal}{2010}]{Smith10}	
Smith G. P., Khosroshahi H. G., Dariush A., Sanderson A. J. R., Ponman T. J., Stott J. P., Haines C. P., Egami E., Stark D. P., 2010,
MNRAS, 409, 169
\bibitem[\protect\citeauthoryear{Spitzer}{1962}]{Spitzer62} 
Spitzer L., 1962, Physics of Fully Ionized Gases. Wiley Interscience, New York, NY
\bibitem[\protect\citeauthoryear{Stawartz, Kneiske \& Kataoka}{2006}]{Stawartz06}
Stawartz L., Kneiske T. M., Kataoka J., 2006, APJ, 637, 693
\bibitem[\protect\citeauthoryear{Stawarz \etal}{2008}]{Stawarz08}
Stawarz Ł., Ostorero L., Begelman M. C., Moderski R., Kataoka J., Wagner S., 2008
 ApJ, 680, 911
\bibitem[\protect\citeauthoryear{Stott \etal}{2012}]{Stott12}
Stott J. P., Hickox R. C., Edge A. C., Collins C. A., Hilton M., Harrison C. D., 
Romer A. K., Rooney P. J., Kay S. T., Miller C. J., Sahlen M., Lloyd-Davies E. J., 
Mehrtens N., Hoyle Ben., Liddle A. R., Viana P. T. P., McCarthy I. G., Schaye Joop, Booth C. M.,
2012, MNRAS, 422, 2213
\bibitem[\protect\citeauthoryear{Sun \etal}{2004}]{Sun04}
Sun M., Forman W., Vikhlinin A., Hornstrup A., Jones C., Murray S. S., 2004
 ApJ, 612, 805
\bibitem[\protect\citeauthoryear{Tavasoli \etal}{2011}]{Tavasoli11}
Tavasoli S., Khosroshahi H. G., Koohpaee A., Rahmani H., Ghanbari J.,
2011, PASP, 123, 1
\bibitem[\protect\citeauthoryear{Tremblay \etal}{2012}]{Tremblay12}	
Tremblay G. R., O'Dea C. P., Baum S. A., Clarke T. E., Sarazin C. L., Bregman J. N., Combes F., Donahue M., Edge A. C., Fabian A. C., Ferland G. J., McNamara B. R., Mittal R., Oonk J. B. R., Quillen A. C., Russell H. R., Sanders J. S., Salome P., Voit G. M., Wilman R. J., Wise M. W., 2012, MNRAS,
424, 1026
\bibitem[\protect\citeauthoryear{Tingay \& Murphy}{2001}]{Tingay01}
Tingay S. J., Murphy D. W., 2001, ApJ, 546, 210
\bibitem[\protect\citeauthoryear{Vikhlinin \etal}{2007}]{Vikhlinin07}
Vikhlinin A., Burenin R., Forman W. R., Jones C., Hornstrup A., Murray S. S., Quintana H., 2007, 
in Heating versus Cooling in Galaxies and Clusters of Galaxies, 48
\bibitem[\protect\citeauthoryear{Vikhlinin \etal}{1999}]{Vikhlinin99}
Vikhlinin A., McNamara B. R., Hornstrup A., Quintana H., Forman W., Jones C., Way M.,
 1999, ApJL, 520, 1
\bibitem[\protect\citeauthoryear{Voevodkin \etal}{2010}]{Voevodkin10}
Voevodkin A., Borozdin K., Heitmann K., Habib S., Vikhlinin A., Mescheryakov A., Hornstrup A., Burenin R.,
 2010, ApJ, 708, 1376
\bibitem[\protect\citeauthoryear{Voit}{2011}]{Voit11}
Voit G. M., 2011, ApJ, 740, 28 
\bibitem[\protect\citeauthoryear{Wadadekar}{2004}]{Wadadekar04}
Wadadekar Y., 2004, A\&A, 416, 35 
\bibitem[\protect\citeauthoryear{Yoshioka \etal}{2004}]{Yoshioka04}
Yoshioka T., Furuzawa A., Takahashi S., Tawara Y., Sato S., Yamashita K., Kumai Y., 2004, AdSpR, 34, 2525
\bibitem[\protect\citeauthoryear{Zakamska \& Narayan}{2003}]{Zakamska03}
Zakamska N. L., Narayan R., 2003, ApJ, 582, 162

\end{thebibliography}
\end{document}